\documentclass[preprint, aip, jmp]{revtex4-1}
\usepackage{natbib}
\usepackage{epsf,graphicx,graphics}
\usepackage{amsmath}
\usepackage{latexsym,amssymb,amsbsy}
\usepackage{enumerate}
\usepackage{setspace}
\usepackage[left=2.54cm,right=2.54cm,top=2.54cm,bottom=2.54cm]{geometry}

\newtheorem{thr}{Theorem}

\newtheorem{lem}[thr]{Lemma}
\newtheorem{prop}[thr]{Proposition}
\newtheorem{cor}[thr]{Corollary}

\usepackage{cleveref}

\usepackage{mathtools}

\newcommand{\rar}{\rightarrow}

\newcommand\rrar{%
        \mathrel{\vcenter{\mathsurround0pt
                \ialign{##\crcr
                        \noalign{\nointerlineskip}$\rightarrow$\crcr
                        \noalign{\nointerlineskip}$\rightarrow$\crcr
                }%
        }}%
}

\newcommand{\slim}{\sum\limits}

\newcommand{\ul}[1]{\underline{#1}}

\newcommand{\disfrac}[2]{\displaystyle{\frac{#1}{#2}}}

\newcommand{\essu}{\mathfrak{su}(2)}
\newcommand{\sun}{\mathfrak{su}(N)}
\newcommand{\suinf}{\mathfrak{su}(\infty)}

\newcommand{\bv}[1]{\mbox{\textbf{#1}}}
\newcommand{\bsym}[1]{\boldsymbol{#1}}

\newcommand{\nat}{\mathbb{N}}

\newcommand{\mc}[1]{\mathcal{#1}}

\newcommand{\mk}[1]{\mathfrak{#1}}

\newcommand{\kron}{\mbox{\boldmath{$\delta$}}}

\newcommand{\newpar}{\\ \\}

\newcommand{\vsp}{\vspace{0.5cm}}

\begin{document}


\title{{Existence of topological hairy dyons and dyonic black holes in anti-de Sitter $\mathfrak{su}(N)$ Einstein-Yang-Mills theory}} 




\author{J. Erik Baxter}
\noaffiliation
\address{Department of Engineering and Mathematics, Sheffield Hallam University, Sheffield, S11WB}
\email[Please address any correspondence to: ]{e.baxter@shu.ac.uk}


\date{\today}

\begin{abstract}
We investigate dyonic black hole and dyon solutions of four-dimensional $\mathfrak{su}(N)$ Einstein-Yang-Mills theory with a negative cosmological constant. We derive a set of field equations in this case, and prove the existence of non-trivial solutions to these equations for any integer $N$, with $2N-2$ gauge degrees of freedom. We do this by showing that solutions exist locally at infinity, and at the event horizon for black holes and the origin for solitons. We then prove that we can patch these solutions together regularly into global solutions that can be integrated arbitrarily far into the asymptotic regime. Our main result is to show that dyonic solutions exist in open sets in the parameter space, and hence that we can find non-trivial dyonic solutions in a number of regimes whose magnetic gauge fields have no zeroes, which is likely important to the stability of the solutions. 
\end{abstract}

\pacs{04.20.Jb, 04.40.Nr, 04.70.Bw}

\maketitle 

\section{Introduction}

Black hole hair has been a topic of discussion in the literature for over forty years now ever since Israel and Carter's original uniqueness (``No-hair") theorems for static, asymptotically flat black holes \cite{WI1967_Schwa_uniq, WI1968_RN_uniq, IH1987_rel_mod_phys} which classified all black holes by their mass $m$ and electric charge $e$. This became a subject of active research in the 1980's due to the discovery by Bartnik and McKinnon \cite{BMcK1988_solitons} and Bizon \cite{PB1990_su2_BH} that these theorems could be violated by considering Einstein-Yang-Mills (EYM) theory, which imposes extra gauge symmetries on space-time -- these manifest as gauge fields coupled to the gravitational field. The spirit of the conjecture was preserved in a way, since the known solutions were still specified by a relatively small number of parameters, and for asymptotically flat space the solutions were found to be sparse, existing with only discrete values of boundary conditions \cite{BFM1994_static_spher_symm, smoller_existence_1993, smoller_existence_1993-1, smoller_smooth_1991}. In addition these are unstable, since it is known that the number of unstable modes of these solutions is proportional to the number of zeroes of the associated gauge field function (``nodes''), with $2n$ unstable modes for gauge functions with $n$ nodes, and in addition, that solutions with $\Lambda=0$ have at least one node \cite{LM1995_inst_BMcK, VBLS1995_sphal_inst_BMcK, MW1996_stab_cons_EYM}. Much work exists in the literature for hairy black holes in the case $\Lambda=0$ involving a wide variety of special cases, including non-spherically symmetric geometries and higher-dimensional analogues \cite{BFM1994_static_spher_symm, lemos_cylindrical_1995, kleihaus_su3_1995, kleihaus_su3_1995-1}. 

The differing geometry bestowed by a negative cosmological constant yields interesting results. For one, we find that the solution space is a lot more rich and abundant, with solutions existing in open sets of the initial conditions as opposed to discrete families \cite{EW1999_su2_ex, BH2000_adS_dyon, BH2000_dyon_BH_4D_EYM, BLM2004_grav_sol_adS}; and we may find `nodeless' solutions (i.e. which possess gauge field functions with no zeroes), which we have said are of importance to stability. When we take $\Lambda<0$, we find solutions where the gauge field function $\omega$ has no zeroes and which are stable under linear \cite{EW1999_su2_ex, BH2000_adS_dyon, BH2000_dyon_BH_4D_EYM, BLM2004_grav_sol_adS} and non-spherically symmetric \cite{winstanley_linear_2002, sarbach_linear_2001} perturbations, provided we take the limit $|\Lambda|\rar\infty$.

The case of hairy particle-like or `soliton' solutions has also been considered. The non-existence of gravitational \cite{heusler_no-hair_1996} or pure Yang-Mills \cite{deser_absence_1976} solitons in asymptotically flat space preceded the surprise discovery of EYM solitons in $\essu$ asymptotically anti-de Sitter (adS) space \cite{BMcK1988_solitons}. We note that these are only stable for $\Lambda<0$ \cite{LM1995_inst_BMcK}, so naturally this inspired generalisations of their work in asymptotically adS space \cite{BLM2004_grav_sol_adS}. Indeed, solitons have been a subject of much interest in a range of special cases involving existence and stability \cite{hosler_higher-dimensional_2009, brihaye_higher_2003, zhou_instability_1990, smoller_regular_1995}.

These ideas have been extended in many ways. One obvious idea is to extend the gauge group from $\essu$ into $\sun$, which is something that that has been worked on previously \cite{volkov_gravitating_1999, HPK1991_suN_spher_symm, BW2008_sun_ex, BWH2008_abund_stable, EW1999_su2_ex, BWH2007_sol_BH_adS_sun}. Another extension is to move away from the assumption of spherical symmetry to considering solutions on manifolds of alternative topology \cite{birmingham_topological_1999, lemos_two-dimensional_1994, mann_topological_2009, vanzo_black_1997, cai_plane_1996}. Notable here is the work of van der Bij and Radu \cite{VdBR_su2_top}, who considered manifolds with isometry groups based on foliations of space-time by surfaces of constant Gaussian curvature: these are parameterised by an integer $k=1,0,-1$, the sign of the curvature; and the foliated surfaces of the space have a spherical ($k=1$), planar ($k=0$) or hyperbolic ($k=-1$) isometry group. That work inspired a previous work by the suthor\cite{JEB2014_top_ex}, and these ideas will appear here. We note early on that while black hole solutions are possible for all 3 values of $k$, solitons are only possible for $k=1$, the spherical case: this is because for $k=0,-1$, the Riemann curvature scalar $R$ blows up at the origin, and therefore there is no such thing as a globally regular solution for $k\neq1$ throughout the range $[0,\infty)$.

Another subject of interest has been in so-called \textit{dyonic} solutions, which have a non-zero electric sector of the gauge potential, unlike previous work which has primarily concerned `purely magnetic' solutions. Non-existence of genuinely non-Abelian regular monopoles and dyons (the dyonic analogue of solitons) \cite{footnote1} was proven for flat space $\essu$ EYM \cite{ershov_non-existence_1990}, but then later black hole solutions and stable monopole and dyon solutions were found in asymptotically adS $\essu$ EYM \cite{BH2000_adS_dyon, BH2000_dyon_BH_4D_EYM}. In addition, dyons and dyonic black holes have been found in spaces with axial symmetry \cite{radu_static_2004, radu_static_2002}. Notably, Nolan and Winstanley  \cite{NW2012_dyon} recently proved the existence of dyons and dyonic black holes in four-dimensional $\essu$ EYM theory with $\Lambda<0$. It is our intention to extend this work in four-dimensional dyonic solutions to consider an $\sun$ gauge field, and for topological black holes of the kind considered by van der Bij and Radu \cite{VdBR_su2_top}. 

In this paper, we prove the existence of four-dimensional, topological, dyonic black hole and soliton solutions to $\sun$ EYM theory with $\Lambda<0$. In Section \ref{ansfe}, we review the features of the model in question, and derive the field equations in this case. We also prove the existence in Section \ref{triv} of several trivial solutions to the equations, whose existence has been proven elsewhere. In Section \ref{lexboundsec}, we prove several propositions concerning the local existence and analyticity of regular solutions near the spatial boundaries of the solutions, i.e. near $r=0$, $r=r_h$ and as $r\rightarrow\infty$. Then in Section \ref{gloexsec}, we prove some propositions concerning the global behaviour of solutions: notably, that solutions which start regularly near the event horizon for black holes (or the origin in the case of solitons) can be integrated arbitrarily far out into the asymptotic regime, as long as the metric function $\mu>0$; and that solutions in the asymptotic regime will remain regular as $r\rar\infty$. We use these propositions in Section \ref{gloexsection} to show we may `stitch' locally existing solutions together into global solutions -- this is a fairly standard `shooting' argument that has a well-established usage in the literature \cite{BMcK1988_solitons, JEB2014_top_ex}. Essentially, we prove that solutions to the dyonic field equations exist in open sets. This allows us to use some trivial solutions and some previously established existing solutions to deduce the existence of non-trivial solutions to the field equations in various regimes. We also prove in Section \ref{l0exsec} that we can find solutions in the limit $|\Lambda|\rar\infty$ for arbitrary initial values for one gauge field and initial values in a neighbourhood of zero for the other. We finish in Section \ref{conclu} by presenting our main results and final conclusions.

\section{4D topological $\sun$ Einstein-Yang-Mills theory}\label{ansfe}

In this section we give the action and ans\"{a}tze for topological $\sun$ EYM theory for $\Lambda<0$, give a brief derivation of the dyonic field equations, and list some trivial solutions in this case.

\subsection{Ans\"{a}tze and field equations}

The action we shall use for four-dimensional $\mathfrak{su}(N)$ EYM theory with a negative cosmological constant is
\begin{equation}\label{action}
S_{EYM}=\frac{1}{2}\int d^{4}x\sqrt{-\mbox{det}g}[R-2\Lambda-\mbox{Tr}F_{\mu\nu}F^{\mu\nu}].
\end{equation}
Here, $R$ is the Ricci scalar of the geometry; we take $\Lambda$, the cosmological constant, to be less than zero; and the anti-symmetric field strength tensor $F_{\mu\nu}$ is given by
\begin{equation}
F_{\mu\nu}=\partial_\mu A_\nu-\partial_\nu A_\mu+q\left[A_\mu,A_\nu\right].
\end{equation} 

In all that follows we use units in which $4\pi G=c=q=1$, and the 4D metric we use has signature $(-, +, +, +)$. Varying the action \eqref{action} gives the field equations
\begin{equation}\label{genFEs}
\begin{split}
T_{\mu\nu}&=R_{\mu\nu}-\frac{1}{2}g_{\mu\nu}R + \Lambda g_{\mu\nu},\\
0&=\nabla_\lambda F^\lambda_{\:\:\mu}+[A_\lambda,F^\lambda_{\:\:\mu}],\\
\end{split}
\end{equation}
where the Yang-Mills stress-energy tensor is
\begin{equation}\label{YMset}
T_{\mu\nu}=\mbox{Tr}F_{\mu\lambda}F_\nu^\lambda-\frac{1}{4}g_{\mu\nu}\mbox{Tr}F_{\lambda\sigma}F^{\lambda\sigma}.
\end{equation}
Note that the square bracket in \eqref{genFEs} is the Lie algebra commutator, and `Tr' in (\ref{action}, \ref{YMset}) is the Lie algebra trace. Throughout this work we have employed the usual Einstein summation convention, unless otherwise stated.

In this paper we focus on static, topological black hole and soliton solutions of the field equations \eqref{genFEs}, specifically for spaces regularly foliated by 2D (spacelike) hypersurfaces of constant Gaussian curvature which can be indexed by $k\in\{-1,0,1\}$ for positive, zero, and negatively curved spaces \cite{JEB2014_top_ex}. Hence, we may write the metric in standard Schwarzschild co-ordinates as
\begin{equation}\label{metric}
ds^2=-\mu S^2dt^2+\mu^{-1}dr^2+r^2d\Omega_k^2
\end{equation}
where $\mu$ and $S$ depend on $r$ alone. We may write the function $\mu$ as
\begin{equation}
\mu(r)=k-\frac{2m(r)}{r}+\frac{r^2}{\ell^2},
\end{equation}
noting that the constant $\ell$ is the adS radius of curvature defined by 
\begin{equation}
\ell^2=-\frac{3}{\Lambda},
\end{equation}
possible because we are only interested in $\Lambda<0$. We will refer to both forms of the cosmological constant in this work. The angular part of the metric, $d\Omega_k^2$, is given by
\begin{equation}
d\Omega_k^2=d\theta^2+f^2_k(\theta)d\phi^2;
\end{equation}
where we recall that if we let
\begin{equation}\label{fkdef}
f_k(\theta) = \left\{ \begin{array}{ll}
\sin\theta & \mbox{for } k=1 \\
\theta & \mbox{for } k=0 \\
\sinh\theta & \mbox{for } k=-1 \end{array}\right.,
\end{equation}
then our metric \eqref{metric} is endowed with the topology appropriate to each case.

In previous work \cite{JEB2014_top_ex} we only considered `purely magnetic' solutions for topological $\sun$ black holes. Here, we wish to include the electric part as well, meaning that we use the full gauge potential, which in our case is
\begin{equation}\label{GP}
A=A_\mu dx^\mu=\mathcal{A}\,dt+\mc{B}\,dr+\frac{1}{2}(C-C^\dagger)d\theta-\frac{i}{2}\left[(C+C^\dagger)f_k(\theta)+D\frac{df_k(\theta)}{d\theta}\right]d\phi,
\end{equation}
where $\mathcal{A},\mc{B},C$ and $D$ are all ($N\times N$) matrices, $C^\dagger$ is the Hermitian conjugate of $C$, and $f_k(\theta)$ is given above as \eqref{fkdef}. The matrices $\mathcal{A}$ and $\mathcal{B}$ are given by
\begin{equation}
[\mathcal{A}]_{jj}=\frac{i}{2}\alpha_{j}(r), \quad [\mathcal{B}]_{jj}=\frac{i}{2}B_{j}(r)
\end{equation}
for $2N$ functions $\alpha_j(r)$ and $B_j(r)$ ($j=1,...,N$) constrained by
\begin{equation}\label{alphacon}
\sum^{N}_{j=1}\alpha_j(r)=\slim_{j=1}^NB_j(r)=0.
\end{equation}
Hence these represent $2N-2$ independent functions. The matrix $C$ (which also depends solely on $r$) is upper-triangular, with non-zero entries only immediately above the diagonal, i.e.:
\begin{equation}\label{Cdef}
[C]_{j,j+1}=\left([C^\dagger]_{j+1,j}\right)^\star=\omega_j(r)e^{ig_j(r)}
\end{equation}
(where $^\star$ is the complex conjugate) for $2N-2$ functions $\omega_j(r)$ and $g_j(r)$ ($j=1,...,N-1$). The matrix $D$ is constant and diagonal:
\begin{equation}\label{Ddef}
[D]_{jj}=N+1-2j.
\end{equation}
We find that one of the Yang-Mills equations gives us $B_j+g_j=0$, hence we let $g_j=-B_j$. Finally we may exploit a remaining gauge freedom to set $B_j=g_j=0$. Therefore we have $2N-2$ independent gauge field functions altogether: $\alpha_j(r)$, which we shall call the \textit{Electric Gauge Functions} (EGFs), and $\omega_j(r)$, which we shall call the \textit{Magnetic Gauge Functions} (MGFs). For further details on this choice of potential, note that the precise form has been derived in detail \cite{JEB2014_top_ex} following a method due to K\"{u}nzle \cite{HPK1991_suN_spher_symm}.

Substituting the metric \eqref{metric} and the gauge potential \eqref{GP} into the field equations \eqref{genFEs} (and using the Bianchi identities) gives us the two Einstein equations
\begin{align}
m'&=\disfrac{r^2\eta}{4S^2}+\disfrac{\zeta}{4\mu S^2}+\mu G+P,\label{EE1}\\
\disfrac{S'}{S}&=\disfrac{\zeta}{2\mu^2 S^2r}+\disfrac{2G}{r},\label{EE2}
\end{align}
where for convenience we have defined new quantities
\begin{align}
\eta&\equiv\sum^N_{j=1}\alpha_j'^2,\label{eta}\\
\zeta&\equiv\sum_{j=1}^{N-1}\omega_j^2\left(\alpha_j-\alpha_{j+1}\right)^2,\label{zeta}\\
G&\equiv\sum_{j=1}^{N-1}\omega_j'^2,\label{G}\\
P&\equiv\frac{1}{4r^2}\sum_{j=1}^{N}\left(\omega_j^2-\omega_{j-1}^2-k(N+1-2j)\right)^2;\label{P}
\end{align}
and $2N-2$ independent Yang-Mills equations
\begin{align}
\alpha_j^{\prime\prime}&=\left(\disfrac{S^\prime}{S}-\disfrac{2}{r}\right)\alpha_j^\prime+\disfrac{1}{{\mu r^2}}\left(\omega_j^2\left(\alpha_j-\alpha_{j+1}\right)-\omega_{j-1}^2\left(\alpha_{j-1}-\alpha_{j}\right)\right),\label{YM1}\\
\omega_j^{\prime\prime}&=-\left(\disfrac{S^\prime}{S}+\disfrac{\mu^\prime}{\mu}\right)\omega_j^\prime-\disfrac{1}{4\mu^2 S^2}\omega_j\left(\alpha_j-\alpha_{j+1}\right)^2-\disfrac{W_j\omega_j}{\mu r^2},\label{YM2}
\end{align}
where we have defined 
\begin{equation}\label{Wj}
\begin{split}
W_j&\equiv k-\omega_j^2+\frac{1}{2}\left(\omega_{j+1}^2+\omega_{j-1}^2\right).\\
\end{split}
\end{equation}

Note that a prime $^\prime$ stands for $d/dr$, and we define $\alpha_0\equiv\alpha_{N+1}\equiv\omega_0\equiv\omega_N\equiv 0$. Also, it may be noted that there are $N$ equations in $\alpha_j$, but only $N-1$ of the equations are independent due to the degree of freedom given to us by \eqref{alphacon}. If we take the above field equations (\ref{EE1}, \ref{EE2}, \ref{YM1}, \ref{YM2}) and let $N=2$ and $k=1$, we can verify that they reduce to the correct limit for dyonic spherically symmetric $\essu$ EYM field equations \cite{NW2012_dyon}; and if we let $\alpha_j(r)\equiv0$ we recover purely magnetic topological $\sun$ EYM equations \cite{JEB2014_top_ex}.

As may be expected from previous cases \cite{EW1999_su2_ex, BW2008_sun_ex, JEB2014_top_ex}, the field equations (\ref{EE1}, \ref{EE2}, \ref{YM1}, \ref{YM2}) are invariant under various transformations. They are invariant under the following two transformations separately:
\begin{equation}
\alpha_j\mapsto-\alpha_j\,\,(\forall j), \quad j\mapsto N-j\,\,(\forall j);
\end{equation}
they are invariant under 
\begin{equation}
\omega_j\mapsto-\omega_j,
\end{equation}	
for \textit{each} $\omega_j$ separately; and finally they are also invariant under the simultaneous transformations
\begin{equation}
\begin{array}{cc}
S(r)\mapsto\lambda S(r),&\quad\quad \alpha_j(r)\mapsto\lambda\alpha_j(r),
\end{array}
\end{equation}
which correspond to applying a time-rescaling $t\mapsto\lambda^{-1}t$ to the metric \eqref{metric} and the potential \eqref{GP}.

Finally, we re-express the equations in another form which will prove useful later. We make the variable change 
\begin{equation}\label{Etrans}
\mc{E}_j=\alpha_j-\alpha_{j+1}.
\end{equation}
This leaves the forms of (\ref{EE1}, \ref{EE2}) unchanged, though it does alter $\zeta$ \eqref{zeta}:
\begin{align}\label{Ezeta}
\zeta\equiv\slim_{j=1}^{N-1}\omega_j^2\mc{E}_j^2.
\end{align}
It also alters $\eta$ but we deal with that as it comes up as the transformation is quite complicated. The Yang-Mills equations are altered to become
\begin{align}
\mc{E}_j^{\prime\prime}&=\left(\disfrac{S^\prime}{S}-\disfrac{2}{r}\right)\mc{E}_j^\prime+\disfrac{1}{{\mu r^2}}\mc{Z}
_{j},\label{YM1E}\\
\omega_j^{\prime\prime}&=-\left(\disfrac{S^\prime}{S}-\disfrac{\mu^\prime}{\mu}\right)\omega_j^\prime-\disfrac{1}{4\mu^2 S^2}\omega_j\mc{E}_j^2-\disfrac{W_j\omega_j}{\mu r^2},\label{YM2E}
\end{align}
with
\begin{equation}\label{Omegaj}
\mc{Z}_j\equiv 2\omega_j^2\mc{E}_j-\omega_{j-1}^2\mc{E}_{j-1}-\omega_{j+1}^2\mc{E}_{j+1}.
\end{equation}
We should mention that since \eqref{Etrans} is an elementary linear transform, it turns out to have an elementary inverse:
\begin{equation}\label{Etransinv}
\alpha_j=\frac{-1}{N}\slim_{k=1}^{j-1}\mc{E}_k+\slim_{k=j}^{N-1}\left(1-\frac{k}{N}\right)\mc{E}_k.
\end{equation}
It can be noticed that this equation has been stated before \cite{HPK1991_suN_spher_symm} as a way of expressing the functions $\alpha_j$, and this connection is to do with the fact that, in vector form, we can express the transform \eqref{Etrans} as $\ul{\mc{E}}=\mc{T}\ul{\alpha}$, where $\mc{T}$ is the $(N-1)\times N$ matrix
\begin{equation}
\mc{T}=\begin{pmatrix}
1 & -1 & 0 & 0 & \hdots & 0 & 0\\
0 & 1 & -1 & 0 & \hdots & 0 & 0\\
0 & 0 & 1 & -1 & \hdots & 0 & 0\\
\vdots & \vdots & \vdots & \vdots & \ddots & \vdots & \vdots\\
0 & 0 & 0 & 0 & \hdots & 1 & -1\\
\end{pmatrix}
\end{equation}
whose rows represent a choice of simple roots for $\sun$ \cite{hall_lie_2003}, and expressing the equations thus will reveal the structure of the Cartan matrix of $\sun$ in the electric gauge equations -- as we shall see, the form of \eqref{Omegaj} simplifies the analysis considerably. 

We lastly note that we can also use the Einstein equations to express \eqref{YM2} in the following useful form:
\begin{equation}\label{YM2E2}
r^2\mu\omega_j^{\prime\prime}=-2\left(m-rP+\frac{r^3}{\ell^2}-\frac{r^3\eta}{4S^2}\right)\omega_j^\prime-\disfrac{r^2}{4\mu S^2}\omega_j\mc{E}_j^2-W_j\omega_j.\\
\end{equation}

\subsection{Trivial solutions}
\label{triv}

We may find the following list of trivial solutions, the existence of most of which has been investigated in previous work.

\begin{enumerate}
\item If we set $\alpha_j(r)\equiv 0$, we recover the purely magnetic topological solutions which we previously investigated \cite{JEB2014_top_ex} (thus they exist for all $k$). We note that in this case, equation (\ref{EE2}) decouples from the others. 
\item If we set $\alpha_j(r)\equiv 0$ and $\omega_j(r)\equiv\sqrt{j(N-j)}$, we obtain the Schwarzschild-anti-de Sitter (SadS) solution where $m(r)=M$ and $S(r)$ is a constant which we usually scale to 1. (This solution only exists for $k=1$.) If $M=0$, we recover pure adS space as a solution.
\item If we set $\alpha_j(r)\equiv\omega_j(r)\equiv 0$, we obtain the magnetically charged Reissner-N\"{o}rdstrom anti-de Sitter (RNadS) black hole solution. Here, $S(r)$ is again a constant but
\begin{equation}
m(r)=M-\disfrac{Q_M}{2r},
\end{equation}
and the magnetic charge $Q_M$ is given by 
\begin{equation}\label{QM}
Q_M=\frac{k^2N(N+1)(N-1)}{6}.
\end{equation}
This solution exists for all $k$.
\item If we instead set $\omega_j(r)\equiv 0$ and $\alpha_j(r)\equiv\disfrac{a_j}{r}$ (with constants $a_j$), we get (for all $k$) an $\sun$ analogy to the $\essu$ Abelian RNadS black hole \cite{NW2012_dyon} with $S(r)=1$,
\begin{equation}
m(r) = M-\disfrac{Q_E+Q_M}{2r};
\end{equation}
where magnetic charge $Q_M$ is given in \eqref{QM}, and total electric charge $Q_E$ is
\begin{equation}
Q_E=\disfrac{1}{2}\sum^N_{j=1}a_j^2,
\end{equation}
in which the individual charges $a_j$ are constrained by
\begin{equation}\label{ajcon}
\quad\sum^N_{j=1}a_j=0.
\end{equation}
We note that apart from \eqref{ajcon} there appear to be no \textit{a priori} constraints on the individual $a_j$s, though one possible choice is $a_j=Q(N+1-2j)$ for some constant $Q$: this represents the $\essu$ embedding of the solution (see \eqref{embed}). Interestingly, this choice would give us
\begin{equation}
Q_E=\disfrac{Q^2N(N-1)(N+1)}{6},
\end{equation}
so that $Q^2Q_M=k^2Q_E$. The reason this is of interest is that for $k=0$ we can evidently find solutions with a \textit{zero} magnetic charge and \textit{non-zero} electric charge, which we know is impossible in the spherical case \cite{NW2012_dyon}. Due to the fact that $\omega_j\equiv0$, we might expect that this solution would not be stable under linear perturbations (in analogy to the $\essu$ results \cite{LM1995_inst_BMcK, VBLS1995_sphal_inst_BMcK, MW1996_stab_cons_EYM, EW1999_su2_ex}); though very recent results we have obtained suggest that for $k\neq 1$ this may not be the case after all \cite{JEB_WIP_stab_top}. Finally, we note that this solution has \textit{not} yet been investigated, though it is worth noting for future reference.

\end{enumerate}

Finally, we can also obtain dyonic $\essu$ embedded solutions with the following rescaling:
\begin{prop}\label{embedprop}
Any solution which satisfies the $\mathfrak{su}(2)$ dyonic field equations \cite{NW2012_dyon} can be rescaled and embedded as an $\mathfrak{su}(N)$ dyonic EYM solution (which satisfies (\ref{EE1}, \ref{EE2}, \ref{YM1}, \ref{YM2})).
\end{prop}
\vsp
\textbf{Proof} We begin with the field equations (\ref{EE1}, \ref{EE2}, \ref{YM1}, \ref{YM2}). We rescale them with the following definitions:
\begin{equation}\label{embed}
\begin{array}{lllll}
\lambda_{N}^2\equiv\frac{1}{6}N(N-1)(N+1), & \quad & r\equiv\lambda_{N}R, & \quad &  \ell\equiv\lambda_{N}\tilde{\ell}, \\
\omega_j\mapsto\sqrt{j(N-j)}\omega\quad\forall j, & \quad & m\equiv\lambda_{N}\tilde{m}, & \quad & \\
\alpha_j\mapsto (N+1-2j)\alpha\quad\!\!\!\forall j, & \quad & S\equiv\lambda_{N}\tilde{S}. & \quad & \\
\end{array}
\end{equation}
This rescaling leads to the following equations:
\begin{equation}
\begin{array}{rl}\label{su2embed}
\disfrac{\mbox{d}\tilde{m}}{\mbox{d}R}&=\disfrac{R^2}{2\tilde{S}^2}\left(\disfrac{\mbox{d}\alpha}{\mbox{d}R}\right)^2+\disfrac{\omega^2\alpha^2}{\mu \tilde{S}^2}+\mu\left(\disfrac{\mbox{d}\omega}{\mbox{d}R}\right)^2+\disfrac{(\omega^2-k)^2}{2R^2},\\
&\\
\disfrac{1}{\tilde{S}}\disfrac{\mbox{d}\tilde{S}}{\mbox{d}R}&=\disfrac{2\omega^2\alpha^2}{\mu^2\tilde{S}^2 R}+\disfrac{2}{R}\left(\disfrac{\mbox{d}\omega}{\mbox{d}R}\right)^2,\\
&\\
\disfrac{\mbox{d}^2\alpha}{\mbox{d}R^2}&=\left(\disfrac{1}{\tilde{S}}\disfrac{\mbox{d}\tilde{S}}{\mbox{d}R}-\disfrac{2}{R}\right)\disfrac{\mbox{d}\alpha}{\mbox{d}R}+\frac{2\omega^2\alpha}{\mu R^2},\\
&\\
\disfrac{\mbox{d}^2\omega}{\mbox{d}R^2}&=-\left(\disfrac{1}{\mu}\disfrac{\mbox{d}\mu}{\mbox{d}R}+\disfrac{1}{\tilde{S}}\disfrac{\mbox{d}\tilde{S}}{\mbox{d}R}\right)\disfrac{\mbox{d}\omega}{\mbox{d}R}-\disfrac{\alpha^2\omega}{\mu^2 \tilde{S}^2}-\disfrac{\omega(\omega^2-k)^2}{\mu R^2}.\\
\end{array}
\end{equation}
It can be checked that if we let $k=1$, these are exactly the same equations as the dyonic spherically symmetric $\mathfrak{su}(2)$ field equations for which existence of solutions has been proven \cite{NW2012_dyon}; and if we let $\alpha\equiv 0$ then we recover the original topological $\essu$ equations \cite{VdBR_su2_top}. We also note that the tracelessness condition \eqref{alphacon} constraining the $\alpha_j$ is still satisfied.$\Box$

\vsp
In this paper we investigate the existence of solutions to the field equations (\ref{EE1}, \ref{EE2}, \ref{YM1}, \ref{YM2}) which do not appear in the above list -- i.e. genuine non-trivial solutions.

\subsection{Boundary conditions}

As we are interested in both black hole and soliton solutions, the boundary points we must examine are the origin $r=0$ for solitons, the event horizon $r=r_h$ for black holes, and for both solutions, the limit $r\rar\infty$. The field equations are singular at all of these points, so we shall now assume appropriate power series solutions regular near the boundary in question, substitute them into the field equations, and also use some requirements of physicality (e.g. expected asymptotic behaviour, regularity) to determine the local power series expansions -- in particular, we are interested in how many independent parameters are required to specify any solution.

\subsubsection{Origin $r=0$}
At the origin, we simply use the co-ordinate $r$. The following power series forms are assumed:
\begin{equation}\label{orexpan}
\begin{split}
m(r)&=m_0+m_1r+m_2r^2+O(r^3),\\
S(r)&=S_0+S_1r+S_2r^2+O(r^3),\\
\omega_j(r)&=\omega_{j,0}+\omega_{j,1}r+\omega_{j,2}r^2+O(r^3),\\
\mc{E}_j(r)&=\mc{E}_{j,0}+\mc{E}_{j,1}r+\mc{E}_{j,2}r^2+O(r^3).\\
\end{split}
\end{equation}
We then substitute these expansions into the field equations (\ref{EE1}, \ref{EE2}, \ref{YM2}, \ref{YM1E}). To avoid a singularity in the metric and the field equations, it must be the case that $S_0$ is non-zero,
\begin{equation}\label{r0parcon}
\begin{split}
m_0&=m_1=m_2=S_1=\mc{E}_{j,0}=\omega_{j,1}=0,\mbox{ and}\\
\omega_{j,0}&=\sqrt{j(N-j)},
\end{split}
\end{equation}
for all $j\in\{1,...,N-1\}$. The metric function expansions are not difficult to compute. The Yang-Mills sector is vastly more complicated and involves solving a matrix equation. We shall look into this in much more detail later in Section \ref{psr0con}; for now we shall merely present the results. We define vectors $\bsym{\omega}\equiv(\omega_1,\omega_2,...,\omega_{N-1})^T$ and $\bsym{\mc{E}}\equiv(\mc{E}_1,\mc{E}_2,...,\mc{E}_{N-1})^T$. When we substitute the conditions \eqref{r0parcon} into the field equations, the analysis implies that the expansion is best done in terms of the $N-1$ eigenvectors of the matrix in question \eqref{matA}, so that in order to obtain all the independent parameters, we must expand $\bsym{\omega}$ up to order $r^N$ and $\bsym{\mc{E}}$ up to order $r^{N-1}$. The expansions nearby the origin \eqref{orexpan} thus become (in components)
\begin{equation}\label{origexp}
\begin{split}
m(r)&=m_3r^3+O(r^4),\\
S(r)&=S_0+S_2r^2+O(r^3),\\
\omega_j(x)=&\sqrt{j(N-j)}\left(1+\slim_{k=1}^{N-1}\bar{\beta}_{k}(x)v^j_kr^{k+1}\right)+O(r^{N+1}),\\
\mc{E}_j(x)=&\slim_{k=1}^{N-1}\bar{\theta}_k(x)v^j_kr^{k}+O(r^N).\\
\end{split}
\end{equation}
We will define the vectors $v^j_k$ properly later; just now all we need state is that $\bar{\beta}_{k}(x)$, $\bar{\theta}_{k}(x)$ are determined entirely by the $2N-2$ parameters $\bar{\beta}_{k}(0)\equiv\breve{\beta}_k$ and $\bar{\theta}_{k}(0)\equiv\breve{\theta}_k$; and the constants $m_3$ and $S_2$ are determined entirely by $S_0$, $\alpha_{j,1}$ (or $\mc{E}_{j,1}$) and $\omega_{j,2}$ by the field equations as follows.
\begin{equation}
\begin{split}
m_3&=\disfrac{1}{12S_0}\left(\sum^N_{j=1}\alpha^2_{j,1}+\sum^{N-1}_{j=1}j(N-j)\mc{E}_{j,1}^2\right)\\
&+\disfrac{4}{3}\sum^{N-1}_{j=1}\omega_{j,2}^2+\disfrac{1}{6}\sum^N_{j=1}\left(\sqrt{j(N-j)}\omega_{j,2}-\sqrt{(j-1)(N-j+1)}\omega_{j-1,2}\right)^2,\\
S_2&=\disfrac{1}{4S_0}\sum^{N-1}_{j=1}j(N-j)\mc{E}_{j,1}^2+4S_0\sum^{N-1}_{j=1}\omega_{j,2}^2.\\
\end{split}
\end{equation}

The value of $S_0$ is fixed by the requirement that $S\rar1$ as $r\rar\infty$, therefore these solutions are specified entirely by $2N-2$ parameters.
\subsubsection{Event horizon $r=r_h$}

At the event horizon, the picture is much clearer. Firstly we have $\mu(r_h)=0$. This means we immediately have
\begin{equation}
m_h=\disfrac{kr_h}{2}+\disfrac{r_h^3}{2\ell^2}.
\end{equation}
It also means that (\ref{YM1}) will be singular at $r=r_h$, unless $\alpha_j(r_h)=0$ for all $j$. Substituting these two facts into the field equations (\ref{EE1}, \ref{EE2}, \ref{YM1}, \ref{YM2}) produces the following regular Taylor expansions in the field variables at the event horizon:
\begin{equation}\label{exprh}
\begin{array}{rl}
m(r)&=m_h+m'_h(r-r_h)+O(r-r_h)^2,\\
S(r)&=S_h+S'_h(r-r_h)+O(r-r_h)^2,\\
\alpha_j(r)&=\alpha'_{j,h}(r-r_h)+O(r-r_h)^2,\\
\omega_j(r)&=\omega_{j,h}+\omega'_{j,h}(r-r_h)+O(r-r_h)^2;\\
\end{array}
\end{equation}
where the expressions for $m'_h$, $S'_h$, $\omega'_{j,h}$ are given in terms of $\alpha'_{j,h}$, $\omega_{j,h}$ and $S_h$:
\begin{equation}
\begin{array}{rl}
m'_h&=\disfrac{r_h^2}{4S_h^2}\sum^N_{j=1}\alpha'^2_{j,h}+\disfrac{1}{4r_h^2}\sum^N_{j=1}\left(\omega_{j,h}^2-\omega_{j+1,h}^2-k(N+1-2j)\right)^2,\\
S'_h&=\disfrac{2S_h}{r_h}\sum^{N-1}_{j=1}\omega'^2_{j,h},\\
\omega'_{j,h}&=-\disfrac{\omega_{j,h}}{\mu'_h r_h^2}\left(k-\omega_{j,h}^2+\frac{1}{2}\left(\omega_{j+1,h}^2+\omega_{j-1,h}^2\right)\right),\\
\end{array}
\end{equation}
Therefore $\mu_h$ and $\mu'_h$ are entirely determined, and we also have
\begin{equation}
\mu'_h\equiv\mu'(r_h)=\disfrac{k}{r_h}-\disfrac{2m'_h}{r_h}+\disfrac{3r_h}{\ell^2}>0,
\end{equation}
where the last inequality is due to the requirement of a regular event horizon. Thus, fixing $r_h$ and $\ell$, and fixing $S_h$ with the requirement that $S\rar1$ as $r\rar\infty$, the solutions near the event horizon can be entirely specified by $2N-2$ parameters.\newpar
We should also finally note a weak constraint on $\omega_j$ and $\alpha_j$ near the horizon, given by requiring a non-extremal event horizon: setting $\mu_h=\alpha_{j,h}=0$ in the field equation for $m'$ gives us
\begin{equation}
2m'(r_h)=\disfrac{r^2_h\eta(r_h)}{2S_h^2}+2P(r_h)<k+\disfrac{3r_h^2}{\ell^2},
\end{equation}
as directly related to the tangential pressure \cite{OK2002_gen_gauge_group_stab}. This also allows us to define a minimum event horizon radius for $k=-1$,
\begin{equation}\label{minrhk-1}
r_h^2>\disfrac{\ell^2}{3}\left(2m'_h+1\right),
\end{equation}
where it can be seen that the right-hand side of the inequality is clearly positive. Finally, we note that it gives us a minimum bound on $|\Lambda|$ for $k=-1$, in analogy to the $\essu$ case \cite{VdBR_su2_top}:
\begin{equation}
|\Lambda|>\frac{1}{r_h^2}\left(1+2P(r_h)+\frac{r_h^2\eta(r_h)}{2S_h^2}\right).
\end{equation}

\subsubsection{Infinity}

As $r\rar\infty$, we wish the solution to approach `topological adS' -- that is, we wish $m\rightarrow M$ (some constant) and $S\rightarrow 1$. We also note that in this regime, $\mu(r)\rar\frac{r^2}{\ell^2}$; hence using (\ref{EE2}) shows that $S'(r)\sim O(r^{-5})$. Therefore, the expansions in the asymptotic region are:
\begin{equation}\label{expinf}
\begin{array}{rl}
m(r)&=M+m_1r^{-1}+O(r^{-2}),\\
S(r)&=1+S_4r^{-4}+O(r^{-5}),\\
\alpha_j(r)&=\alpha_{j,\infty}+d_jr^{-1}+O(r^{-2}),\\
\omega_j(r)&=\omega_{j,\infty}+c_jr^{-1}+O(r^{-2}),\\
\end{array}
\end{equation}
where $m_1$ and $S_4$ are given in terms of $\alpha_{j,\infty}$, $d_j$, $\omega_{j,\infty}$ and $c_j$:
\begin{equation}
\begin{array}{rl}
m_1=&-\disfrac{1}{4}\sum^N_{j=1}d_j^2-\disfrac{\ell^2}{4}\sum_{j=1}^{N-1}\omega^2_{j,\infty}\left(\alpha_{j,\infty}-\alpha_{j+1,\infty}\right)^2-\disfrac{1}{\ell^2}\sum_{j=1}^{N-1}c_j^2\\
&-\disfrac{1}{4}\sum^N_{j=1}\left(\omega^2_{j,\infty}-\omega^2_{j+1,\infty}-k(N+1-2j)\right)^2,\\
S_4=&-\disfrac{\ell^4}{8}\sum_{j=1}^N\omega^2_{j,\infty}\left(\alpha_{j,\infty}-\alpha_{j+1,\infty}\right)^2-\disfrac{1}{2}\sum^{N-1}_{j=1}c_j^2.\\
\end{array}
\end{equation}

Note that it is easy to substitute $\alpha_j$ for $\mc{E}_j$ here using \eqref{Etrans} or \eqref{Etransinv}, if we need to obtain the expansions in terms either variable. No conditions are placed on the constants $\alpha_{j,\infty}$, $\omega_{j,\infty}$, $c_j$, $d_j$ or $M$, hence (fixing $r_h$ and $\ell$) we get a $4N-3$ parameter family of solutions.

\section{Local existence of solutions near boundaries}
\label{lexboundsec}

We now begin by proving that solutions to (\ref{EE1}, \ref{EE2}, \ref{YM1}, \ref{YM2}) exist in some neighbourhood of the boundary points $r=0$, $r=r_h$ and $r\rar\infty$, and that these solutions are analytic in their boundary conditions in some sufficiently small open neighbourhood of the parameter space. To do this we use a well-established a theorem of differential equations, which has a rich history of use in the literature \cite{BFM1994_static_spher_symm, NW2012_dyon, OK2002_gen_spher_EYM}. We begin by stating the theorem.

\begin{thr}\cite{coddington_theory_1955}\label{exdetheor}
Consider a system of differential equations for $n+m$ functions $\mathbf{a}=(a_1,a_2,\ldots,a_n)$ and $\mathbf{b}=(b_1,b_2,\ldots,b_m)$ of the form
\begin{equation} \label{theorem}
\begin{split}
x\frac{da_i}{dx}&=x^{p_i}f_i(x,\mathbf{a},\mathbf{b}),\\
x\frac{db_i}{dx}&=-\lambda_i
b_i+x^{q_i}g_i(x,\mathbf{a},\mathbf{b})\\
\end{split}
\end{equation}
with constants $\lambda_i>0$ and integers $p_i,q_i\geq1$ and let $\mathcal{C}$ be an open subset of $\mathbb{R}^n$ such that the functions $f_i$ and $g_i$ are analytic in a neighbourhood of $x=0$, $\mathbf{a}=\mathbf{c}$, $\mathbf{v}=\mathbf{0}$, for all  $\mathbf{c}\in\mathcal{C}$. Then there exists an $n$-parameter family of solutions of the system such that
\begin{equation}\label{bcs}
\begin{matrix}
&a_i(x)=c_i+O(x^{p_i}),&b_i=O(x^{q_i}),
\end{matrix}
\end{equation}
where $a_i(x)$ and $b_i(x)$ are defined for $\mathbf{c}\in\mathcal{C}$, 
$|x|<x_0(\mathbf{c})$ and are analytic in $x$ and $\mathbf{c}$.
\end{thr}

This theorem allows us to parametrise the family of solutions near a singular point of a set of ordinary differential equations. We need to take each boundary point in turn and transform our field variables so that the field equations are in the form required by Theorem \ref{exdetheor}. After that, it is elementary to verify the forms we have chosen for our expansions of the field variables near the singular points (\ref{origexp}, \ref{exprh}, \ref{expinf}).

\subsection{Power series results at $r=0$ ($k=1$ only)}

\label{psr0con}

For the boundaries $r=r_h$ and $r\rightarrow\infty$, we saw that physicality requirements fixed all of the boundary conditions in a relatively simple way. In the case of solitons we anticipate the situation will be much more intricate, as it was for the purely magnetic $\sun$ cases studied for flat space \cite{HPK1994_analysis_sun} and for asymptotically adS space \cite{BW2008_sun_ex}, and computation of lower order terms confirms this assertion. This is due to the complicated intercoupling between the gauge functions. Therefore we now study the boundary conditions at the origin in detail, which are obtained by assuming power series which are good near the origin and requiring that the metric and field equations are regular there. Our strategy at first then is to take the field equations and convert them from differential equations in the field variables into recurrence relations in the series expansion parameters. We then attempt to find a consistent solution to these relations.\newpar
We begin with the field equations in the forms (\ref{EE1}, \ref{EE2}, \ref{YM1E}, \ref{YM2E}), multiplied through by factors of $\mu$ and $S$ as appropriate, anticipating the later series expansion. Note that we are here using $\mu$ in the form 
\begin{equation}
\mu(r)=1-\disfrac{2m(r)}{r}+\disfrac{r^2}{\ell^2}.
\end{equation}
Then we use the $\essu$ embeddings to find a rescaling of all quantities, as follows. First we define the quantities
\begin{equation}
\begin{array}{lcl}
\gamma_j\equiv j(N-j) & \quad &\mbox{(for }j\in\{1,2,...,N-1\}),\\
&&\\
\lambda_N^2\equiv\disfrac{N(N^2-1)}{6}, & \quad & \kappa_N\equiv\lambda_N^{-2},\\
\end{array}
\end{equation}
and then we rescale:
\begin{equation}\label{r0rescale}
\begin{array}{lclcl}
r=\lambda_Nx, & \quad & m(r)=\lambda_N\tilde{m}(x), & \quad & \ell=\lambda_N\tilde{\ell},\\
&&&&\\
\omega_j(r)\equiv\gamma_j^{\frac{1}{2}}u_j(\lambda_Nx), & \quad & \mc{E}_j(r)=\tilde{\mc{E}}_j(\lambda_Nx) & \quad & \mbox{ (for }j\in\{1,...,N-1\}),\\
&&&&\\
\mu(r)=\tilde{\mu}(\lambda_Nx), & \quad & S(r)=S_0\tilde{S}(\lambda_Nx). & &\\
\end{array}
\end{equation}
%
Finally, for ease of computation we define the quantity
\begin{equation}\label{qj}
q_j=\gamma_ju_j^2-\gamma_{j-1}u_{j-1}^2-N-1+2j.
\end{equation}
All of this brings the field equations into the forms
\begin{equation}\label{FEr02}
\begin{split}
\left(\tilde{\mu}\tilde{S}^2\right)\frac{d\tilde{m}}{dx}=&\left(\tilde{\mu}\tilde{S}^2\right)\kappa_N\left(\tilde{\mu}\tilde{G}+\tilde{P}\right)+\frac{\tilde{\zeta}}{4S_0^2}+\frac{x^2\tilde{\eta}}{4S_0^2},\\
x\left(\tilde{\mu}^2\tilde{S}\right)\frac{d\tilde{S}}{dx}=&2\kappa_N\tilde{G}\tilde{S}\left(\tilde{\mu}^2\tilde{S}\right)+\frac{\tilde{\zeta}}{2S_0^2},\\
x^2\tilde{\mu}\tilde{S}\frac{d^2\tilde{\mc{E}}_j}{dx^2}=&\tilde{\mu}\left(x\frac{d\tilde{S}}{dx}-2\tilde{S}\right)x\frac{d\tilde{\mc{E}}_j}{dx}\\
&+\tilde{S}\left(2\gamma_ju_{j}^2\tilde{\mc{E}}_j-\gamma_{j-1}u_{j-1}^2\tilde{\mc{E}}_{j-1}-\gamma_{j+1}u_{j+1}^2\tilde{\mc{E}}_{j+1}\right),\\
x^2\tilde{\mu}\left(\tilde{\mu}\tilde{S}^2\right)\frac{d^2u_j}{dx^2}=&-2\left(\tilde{\mu}\tilde{S}^2\right)\left(\tilde{m}-\kappa_Nx\tilde{P}+\frac{x^3}{\tilde{\ell}^2}\right)\frac{du_j}{dx}+\frac{x^3\tilde{\mu}\tilde{\eta}}{2S_0^2}\frac{du_j}{dx}\\
&-\frac{\tilde{\mu}\tilde{S}^2}{2}\left(q_{j+1}-q_j\right)u_j-\frac{\lambda_N^2x^2}{4S_0^2}u_j\tilde{\mc{E}}_j^2,\\
\end{split}
\end{equation}
with
\begin{equation}
\begin{array}{ll}
G=\kappa_N\slim_{j=1}^{N-1}\gamma_j\left(\disfrac{du_j}{dx}\right)^2\equiv\kappa_N\tilde{G}, & \quad P=\disfrac{\kappa_N}{4x^2}\slim_{j=1}^Nq_j^2\equiv\kappa_N\tilde{P},\\
\eta=\kappa_N\slim_{j=1}^{N}\left(\disfrac{d\alpha_j}{dx}\right)^2\equiv\kappa_N\tilde{\eta}, & \quad \zeta=\slim_{j=1}^{N-1}\gamma_ju_j^2\tilde{\mc{E}}_j^2.\\
\end{array}
\end{equation}
Requiring that $\tilde{\mu}$, $\frac{d\tilde{\mu}}{dx}$ and the field equations themselves are regular, and noting our rescaling of $S$, leads to the following requirements for the lower order terms:
\begin{equation}
\tilde{S}_0=1, \quad u_{j,0}=1, \quad \tilde{m}_0=\tilde{m}_1=\tilde{m}_2=\tilde{S}_1=u_{j,1}=\tilde{\mc{E}}_{j,0}=0.
\end{equation}
(For clarity, we'll now drop tildes.) This means the power series will have the basic forms
\begin{equation}\label{r0powser}
\begin{array}{lcl}
m(x)=\sum\limits_{k=3}^\infty m_k x^k, & \quad & S(x)=1+\sum\limits_{k=2}^\infty S_k x^k, \\ 
&\\
u_j(x)=1+\sum\limits_{k=2}^\infty u_{j,k} x^k, & \quad & \mc{E}_j(x)=\sum\limits_{k=1}^\infty \mc{E}_{j,k} x^k. \\
\end{array}
\end{equation}
To `sweep up' factors of $\mu$ and $S$ which came from multiplying through originally, we make the replacements
\begin{equation}
\begin{array}{ccc}
\mu S^2\equiv 1 + \hat{M}, & \quad & \mu^2S\equiv 1 + \bar{M},\\
\end{array}
\end{equation}
so that all the summation terms are within $\hat{M}$ and $\bar{M}$, which are given by
\begin{equation}
\begin{split}
\hat{M} \equiv  \sum\limits_{k=2}^\infty \hat{M}_kx^k =& \sum\limits_{k=2}^\infty\left(\mu_k + 2S_k\right)x^k + \sum\limits_{k=4}^\infty\sum\limits_{l=2}^{k-2}\left(S_lS_{k-l} + 2\mu_lS_{k-l}\right)x^k \\
&+\sum\limits_{k=6}^\infty\sum\limits_{l=2}^{k-4}\sum\limits_{p=2}^{k-l-2}\mu_pS_lS_{k-l-p}x^k, \\
\bar{M} \equiv \sum\limits_{k=2}^\infty \bar{M}_kx^k =& \sum\limits_{k=2}^\infty\left(S_k+2\mu_k\right)x^k + \sum\limits_{k=4}^\infty\sum\limits_{l=2}^{k-2}\left(\mu_l\mu_{k-l}+2\mu_lS_{k-l}\right)x^k \\
&+ \sum\limits_{k=6}^\infty\sum\limits_{l=2}^{k-4}\sum\limits_{p=2}^{k-l-2}\mu_p\mu_lS_{k-l-p}x^k. \\
\end{split}
\end{equation}

It can be seen that writing out the field equations using these quantities makes the picture considerably clearer:
\begin{equation}\label{FEr03}
\begin{split}
\frac{dm}{dx}=&\kappa_N\left(\mu G + P\right)+\hat{M}\left(-\frac{dm}{dx}+\kappa_N\left(\mu G + P\right)\right)+\frac{x^2\mu\eta}{4S_0^2}+\frac{\zeta}{4S_0^2},\\
&\\
\frac{dS}{dx}=&\frac{2\kappa_NSG}{x} + \bar{M}\left(-\frac{dS}{dx}+\frac{2\kappa_NSG}{x}\right)+\frac{\zeta}{2S_0^2x},\\
&\\
x^2\mu\frac{d^2u_j}{dx^2}=& -2\left(m-\kappa_nxP+\frac{x^3}{\ell^2}\right)\frac{du_j}{dx} -\frac{1}{2}\left(q_{j+1}-q_j\right)u_j\\
&+\hat{M}\left(-x^2\mu\frac{d^2u_j}{dx^2}-2\left(m-\kappa_nxP+\frac{x^3}{\ell^2}\right)\frac{du_j}{dx}\right.\\
&\left.-\frac{1}{2}\left(q_{j+1}-q_j\right)u_j\right)+\frac{x^3\mu\eta}{2S_0^2}\frac{du_j}{dx}-\frac{x^2\lambda_N^2}{4S_0^2}u_j\mc{E}_j^2,\\
\end{split}
\end{equation}
(and the $\mc{E}_j$ equation in \eqref{FEr02} is unaltered). Thus it may be seen for each of \eqref{FEr03} that the terms divide up into three groups -- terms which have been examined previously \cite{BW2008_sun_ex}, those same terms multiplied through by a factor of $\hat{M}$ or $\bar{M}$, and some extra terms in respect of the non-zero EGFs. This makes the analysis a lot more tractable.

Now we inspect the recurrence relations given by substituting the power series \eqref{r0powser} into the field equations \eqref{FEr03}. We begin with the Einstein equations as they are much easier.

The recurrence relation for $m_k$ is given by:
\begin{equation}\label{recm}
\begin{split}
(k+1)m_{k+1}=&\kappa_N\left(G_k+\frac{1}{\ell^2}G_{k-2}+P_k\right)+\frac{\eta_{k-4}}{4\ell^2S_0^2}+\frac{\zeta_{k}}{4S_0^2}+\frac{\eta_{k-2}}{4S_0^2}\\
&+\sum\limits_{l=2}^{k-2}\left[-2m_{l+1}G_{k-l}-\frac{m_{l+1}\eta_{k-l}}{2S_0^2}\right.\\
&\left.+\hat{M}_l\left(-(k-l+1)m_{k-l+1}+\kappa_N\left(\frac{G_{k-l-2}}{\ell^2}+G_{k-l}+P_{k-l}\right)\right.\right.\\
&\left.\left.-2\sum\limits_{p=3}^{k-l-1}m_pG_{k-l-p+1}\right)\right],\\
\end{split}
\end{equation}
where
\begin{equation}
\begin{split}
\eta_k=&\slim_{j=1}^N\slim_{l=0}^k(l+1)(k-l+1)\alpha_{j,l+1}\alpha_{j,k-l-1},\\
G_k=&\slim_{j=1}^N\slim_{l=1}^{k-1}(l+1)(k-l+1)u_{j,l+1}u_{j,k-l+1},\\
\zeta_k=&\slim_{j=1}^{N-1}\slim_{l=1}^{k-1}\left(\mc{E}_{j,l}\mc{E}_{j,k-l}+2\slim_{p=1}^{k-l-1}\left[u_{j,p}\mc{E}_{j,l}\mc{E}_{j,k-l-p}+\slim_{r=2}^{k-l-p-1}u_{j,r}u_{j,p}\mc{E}_{j,l}\mc{E}_{j,k-l-p-r}\right]\right),\\
P_k=&\frac{1}{4}\slim_{j=1}^N\slim_{l=2}^kq_{j,l}q_{j,k-l+2},\\
q_{j,k}=&2\gamma_ju_{j,k}-2\gamma_{j-1}u_{j-1,k}+\slim_{l=2}^{k-2}\left(\gamma_ju_{j,l}u_{j,k-l}-\gamma_{j-1}u_{j-1,l}u_{j-1,k-l}\right);
\end{split}
\end{equation}
we note that $G_k$, $\zeta_k$, $P_k$ and $q_{j,k}$ are non-zero only for $k\geq 2$ (though $\eta_k\neq 0$ for all $k\geq0$). Note that we leave $\eta_k$ in terms of $\alpha_{j,k}$ -- this is purely for simplicity, as we recognise that $\mc{E}_{j,k}=\alpha_{j,k}-\alpha_{j+1,k}$ for all $k\in\nat$ and thus terms of order $r^k$ in $\alpha_j$ depend only on terms of order $r^k$ in $\mc{E}_j$. Upon examination, we see that $m_{k+1}$ depends only on $m_3,...,m_{k-1}$, $S_0, S_2,...,S_{k-2}$, $u_{j,2},...,u_{j,k}$ and $\mc{E}_{j,1},...,\mc{E}_{j,k-1}$.

The recurrence relation for $S$ is:
\begin{equation}\label{recS}
\begin{split}
kS_k=&2\kappa_NG_k+\frac{\zeta_k}{2S_0^2}+\sum\limits_{l=2}^{k-2}\left(2\kappa_NG_lS_{k-l}+\hat{M}_l\left[-(k-l)S_{k-l}\right.\right.\\
&\left.\left.+2\kappa_N\left(G_{k-l}+\sum\limits_{p=2}^{k-l-2}S_pG_{k-l-p}\right)\right]\right),\\
\end{split}
\end{equation}
and we readily observe that $S_k$ depends only on $m_3,...,m_{k-1}$, $S_0, S_2,...,S_{k-2}$, $u_{j,2},...,u_{j,k}$ and $\mc{E}_{j,1},...,\mc{E}_{j,k-1}$. So as long as we can solve the Yang-Mills recurrence equations to find a consistent regular solution at $r=0$, then we can use those parameters to find $m_k$ and $S_k$ at each order.

Now we come to the much more complicated Yang-Mills equations. They can be expressed as the following:
\begin{equation}\label{YMsys}
\begin{split}
\left[A^i_j-k(k+1)\kron^i_j\right]\bsym{\mc{E}}^j&=\bsym{z}^i,\\
\left[A^i_j-k(k+1)\kron^i_j\right]\bsym{u}^j&=\bsym{b}^i.\\
\end{split}
\end{equation}
We define $\kron^i_j$ as the Kronecker symbol, $A^i_j$ as the $(N-1)\times(N-1)$ matrix given by
\begin{equation}\label{matA}
A^i_j=\left(2\kron^i_j-\kron^{i-1}_j-\kron^{i+1}_j\right)\gamma_j;
\end{equation}
and we have defined $2N-2$ vectors of length $(N-1)$ (in components):
\begin{equation}
\mc{E}^j_k=(\mc{E}_{1,k},\mc{E}_{2,k},...,\mc{E}_{N-1,k})^T, \quad 
u^j_k=(u_{1,k},u_{2,k},...,u_{N-1,k})^T;
\end{equation}
and two length-$(N-1)$ vectors $\bsym{z}^i$ and $\bsym{b}^i$ representing the right hand sides, given by:
\begin{equation}\label{zi}
\begin{split}
z^i_k=&-\sum\limits_{l=1}^{k-2}\left[\mc{E}^i_l\mc{N}_{k,l}+\sum\limits_{j=1}^{N-1}\left[2A^i_ju^j_{l+1}\mc{E}^j_{k-l-1}+\sum\limits_{p=2}^{k-l-2}A^i_ju^j_pu^j_{l+1}\mc{E}^j_{k-l-p-1}\right.\right.\\
&\left.\left.+S_{l+1}\left(\left[A^i_j-(k-l-1)(k-2l-1)\kron^i_j\right]\mc{E}^j_{k-l-1}\right.\right.\right.\\
&\left.\left.\left.+\sum\limits_{p=2}^{k-l-2}\left(2A^i_ju^j_p\mc{E}^j_{k-l-p-1}+\sum\limits_{r=2}^{k-l-p-2}A^i_ju^j_ru^j_p\mc{E}^j_{k-l-p-r-1}\right)\right)\right]\right]\\
\end{split}
\end{equation}
and
\begin{equation}\label{bi}
\begin{split}
b^i_{k+1}=&-\sum\limits_{l=1}^{k-2}\left[u^i_{l+1}\mc{M}_{k,l}+\sum\limits_{j=1}^{N-1}\left(\frac{1}{2}A^i_ju^j_{l+1}u^j_{k-l}+u^i_{k+1}A^i_ju^j_{k-l}\right.\right.\\
&\left.\left.+\frac{1}{2}u^i_{l+1}\sum\limits_{p=2}^{k-l-2}A^i_ju^j_pu^j_{k-l-p}-\hat{M}_{l+1}\left(\left[A^i_j-(k-l)(k-l-1)\kron^i_j\right]u^j_{k-l}\right.\right.\right.\\
&\left.\left.\left.+\sum\limits_{p=2}^{k-l-2}\left[\frac{1}{2}A^i_ju^j_pu^j_{k-l-p}+u^i_pA^i_ju^j_{k-l-p}+\frac{1}{2}u^i_p\sum\limits_{r=2}^{k-l-p-2}A^i_ju^j_ru^j_{k-l-p-r}\right]\right)\right)\right.\\
&\left.+\frac{\lambda_N^2}{4S_0^2}\left(\mc{E}^i_l\mc{E}^i_{k-l-1}+\sum\limits_{p=2}^{k-l-2}u^i_p\mc{E}^i_l\mc{E}^i_{k-l-p-1}\right)\right];\\
\end{split}
\end{equation}
with $\mc{N}_{k,l}$ being the $(N-1)\times(N-1)$ matrix given by
\begin{equation}
\begin{split}
\mc{N}_{k,l}\equiv&2l(l+1)m_{k-l+1}+\frac{1}{\ell^2}\left((k-2l-3)S_{k-l-2}-(k-1)(k-2)\kron^l_{k-2}\right)\\
&-2\sum\limits_{p=2}^{k-l-2}l(p-l-1)S_pm_{k-l-p+1},
\end{split}
\end{equation}
and $\mc{M}_{k,l}$ being the $(N-1)\times(N-1)$ matrix given by
\begin{equation}
\begin{split}
\mc{M}_{k,l}\equiv&(l+1)\left[2(l-1)m_{k-l}-\frac{1}{\ell^2}l\kron^l_{k-2}+2\kappa_NP_{k-l-1}\right.\\
&\left.-\frac{1}{\ell^2}(l+2)\hat{M}_{k-l-3}+\frac{\eta_{k-l-3}}{2S_0^2}+\frac{\eta_{k-l-5}}{2S_0^2\ell^2}\right.\\
&\left.+\sum\limits_{p=2}^{k-l-2}\left[2(l-1)\hat{M}_pm_{k-l-p}-2\kappa_N\hat{M}_pP_{k-l-p-1}-\frac{m_{p+1}\eta_{k-l-p-3}}{4S_0^2}\right]\right].\\
\end{split}
\end{equation}
The system with which we are left thus bears similarities to the purely magnetic system  \cite{HPK1994_analysis_sun}, and so a sensible strategy here is to use results in that work. We expand the matrix $A^i_j$ and the solution vectors $\bsym{u}^i$ and $\bsym{\mc{E}}^i$ in the left- and right-eigenvectors of $A^i_j$, and we attempt to prove that each new vector in the expansions requires only one new parameter each, and hence that the solutions to this equation are consistent in their choice of parameters. Therefore we are in a position to state our proposition:

\begin{prop}\label{conprop}
The system given by the recurrence relations \eqref{recm}, \eqref{recS} and the mutual Yang-Mills system
\begin{equation}\label{YMYM}
\begin{split}
&(A-k(k+1)\bv{I}\,)\bsym{u}_{k+1} = \bsym{b}_{k+1},\\
&(A-k(k+1)\bv{I}\,)\bsym{\mc{E}}_{k} = \bsym{z}_{k},\\
\end{split}
\end{equation}
with conditions
\begin{equation}\label{conscond}
\begin{split}
&\bsym{\sigma}^k\bsym{\mc{E}}_{k} = d_k\theta_k,\\
&\bsym{\sigma}^k\bsym{u}_{k+1} = d_k\beta_k,\\
\end{split}
\end{equation}
has a consistent solution in its parameters, i.e. that all parameters $m_k$, $S_k$, $u_{j,k}$, $\mc{E}_{j,k}$ depend only on $m_3, ..., m_{k-1}$, $S_0, ..., S_{k-1}$, $u_{j,2},...,u_{j,k-1}$ and $\mc{E}_{j,1},...,\mc{E}_{j,k-1}$; thus each new parameter in the expansions can be obtained from those previously calculated. That is to say, each order of the gauge field expansion will contain only one new parameter, for $1\leq k\leq N-1$; and since the metric expansions are determined entirely by physicality and the gauge field parameters, this gives us therefore $2N-2$ solution parameters in total.
\end{prop}
\vsp
\textbf{Proof} We have already stated that if the gauge functions have a consistent solution then the metric functions will too, so we focus on the Yang-Mills system (\ref{YMYM}, \ref{conscond}). The key is that because both systems contain the matrix $A^i_j$, we may appeal to lemmata 1, 2 and 3 formulated by K\"{u}nzle \cite{HPK1994_analysis_sun}, the main results of which we summarise below:

\begin{itemize}
\item  The matrix $A^i_j$ has eigenvalues $1.2$, $2.3$, ... , $(N-1).N$; i.e. the eigenvalues of $A^i_j$ are $k(k+1)$ for $k\in\{1,...,N-1\}$.
\item The left and right eigenvectors of $A^i_j$, $\{\bsym{\sigma}^k\}$ and $\{\bsym{v}_k\}$ respectively, are given by
\begin{equation}\label{evec}
\begin{array}{lll}
\sigma^k_j=\frac{1}{N-1}\gamma_jv^j_k, & \quad & v^j_k=\frac{N-1}{N-j}Q_k(j-1),\\
\end{array}
\end{equation}
where $Q_j(r)$ is defined using Hahn polynomials, with normalisation
\begin{equation}\label{vwnorm}
\slim_{j=1}^{N-1}\sigma^k_jv^j_l=\langle \bsym{v}_k,\bsym{v}_l\rangle = d_k\kron^k_l,
\end{equation}
and normalisation constants $d_k$ given by
\begin{equation}\label{dk}
d_k=\frac{(N+k)!(N-k-1)!}{(N-1)!(N-2)!k(k+1)(2k+1)}.
\end{equation}
(Further details on this may be found in the literature \cite{SK1961_Hahn}.)
\item The system given by
\begin{equation}\label{umat}
\begin{array}{rcl}
(A-k(k+1)\bv{I})\bsym{u}_{k+1} & = & \bsym{b}_{k+1},\\
\bsym{\sigma}^k\bsym{u}_{k+1} & = & d_k\beta_k,\\
\end{array}
\end{equation}
has a consistent solution near $x=0$ regular in all field variables.
\end{itemize}
Examining the $\mc{E}_j$ system in this notation, we find it is very similar to the form of the $u_j$ system:
\begin{equation}\label{emat}
\begin{array}{rcl}
(A-k(k+1)\bv{I})\bsym{\mc{E}}_{k} & = & \bsym{z}_{k},\\
\bsym{\sigma}^k\bsym{\mc{E}}_{k} & = & d_k\theta_k.\\
\end{array}
\end{equation}
Therefore lemmata 1 to 3 will apply equally well to the mutual system (\ref{umat}, \ref{emat}) -- for example, see the proof of local existence near $r\rar\infty$ due to K\"{u}nzle \cite{HPK1994_analysis_sun} -- and so we can follow the basic form of this proof. 

Examining the right-hand sides of (\ref{zi}, \ref{bi}) and find that
\begin{equation}
\bsym{b}_1=\bsym{b}_2=\bsym{z}_1=\bsym{z}_2=0.
\end{equation}
Now, examining lemma 3 in K\"{u}nzle \cite{HPK1994_analysis_sun}, we find
\begin{equation}
\begin{array}{lcl}
\bsym{u}_2=\beta_1\bsym{v}_1, & \quad & \bsym{u}_3=\beta_2\bsym{v}_2,\\
\bsym{\mc{E}}_1=\theta_1\bsym{v}_1, & \quad & \bsym{\mc{E}}_2=\theta_2\bsym{v}_2;\\
\end{array}
\end{equation}
and by a similar argument, we get
\begin{equation}
\mc{E}^i_k=\mc{E}^{i*}_k+\theta_kv^j_k, \quad u^i_{k+1}=u^{i*}_{k+1}+\beta_kv^j_k,\\
\end{equation}
for $3\leq k\leq N-1$. Here, $\mc{E}^{i*}_k$ and $u^{i*}_k$ are special solutions fixed by the requirements
\begin{equation}\label{req}
\bsym{\sigma}^k\cdot\bsym{b}_{k+1}=\bsym{\sigma}^k\cdot\bsym{z}_{k}=0.
\end{equation}
We emphasise that the $\mc{E}^{i*}_k$, $u^{i*}_k$ are entirely specified by \eqref{req} using the results in K\"{u}nzle \cite{HPK1994_analysis_sun}, and then the $2N-2$ arbitrary constants $\beta_k$, $\theta_k$ entirely specify $\mc{E}^i_k$ and $u^i_{k+1}$. 

To perform the analysis it is convenient to expand our equations in terms of the left and right eigenvector basis, and we may use the same basis for both sets of equations according to the above facts. In this basis $A^i_j$ may be written
\begin{equation}\label{matAbasis}
A^i_j=\sum\limits_{a=1}^{N-1}a(a+1)d_a^{-1}v^i_a\sigma^a_j,
\end{equation}
and the gauge field vectors can be expanded in the same basis as
\begin{equation}\label{uebasis}
\begin{array}{lcl}
u^i_{k+1}=\sum\limits_{l=1}^{N-1}U^l_kv^i_l, & \quad & \mc{E}^i_{k}=\sum\limits_{l=1}^{N-1}E^l_kv^i_l,\\
\end{array}
\end{equation}
for constants $U^l_k$ and $E^l_k$. Note also that \eqref{conscond}, \eqref{vwnorm} and \eqref{uebasis} imply that
\begin{equation}\label{Ukk}
U^k_k=\beta_k, \quad E^k_k=\theta_k.
\end{equation}
Finally, for later convenience we define the quantities
\begin{equation}
\begin{array}{lcl}
d^a_{rs}\equiv d_a^{-1}\sum\limits_{i=1}^{N-1}\sigma^a_iv^i_rv^i_s, & \quad & d^a_{rst}\equiv d_a^{-1}\sum\limits_{i=1}^{N-1}\sigma^a_iv^i_rv^i_sv^i_t.
\end{array}
\end{equation}
We may thus express \eqref{YMsys} as
\begin{equation}\label{Eak}
\begin{split}
(a-k)&(a+k+1)E^a_k=-\sum\limits_{l=1}^{k-2}\left[\mc{N}_{k,l}E^a_l+2a(a+1)\sum\limits_{r,s=1}^{N-1}d^a_{rs}U^r_lE^s_{k-l-1}\right.\\
&\left.+\sum\limits_{p=2}^{k-l-2}a(a+1)\sum\limits_{r,s,t,u=1}^{N-1}d^a_{rst}U^r_{p-1}U^s_lE^t_{k-l-p-1}\right.\\
&\left.+S_{l+1}\left(\left[a(a+1)-(k-l-1)(k-2l-1)\right]E^a_{k-l-1}\right.\right.\\
&\left.\left.+\sum\limits_{p=2}^{k-l-2}\left(\sum\limits_{r,s=1}^{N-1}2a(a+1)d^a_{rs}U^r_{p-1}E^s_{k-l-p-1}\right.\right.\right.\\
&\left.\left.\left.+\sum\limits_{r=2}^{k-l-p-2}\sum\limits_{q,s,t,u=1}^{N-1}a(a+1)d^a_{qst}U^q_{r-1}U^s_{p-1}E^t_{k-l-p-r-1}\right)\right)\right],
\end{split}
\end{equation}
\begin{equation}\label{Uak}
\begin{split}
(a-k)&(a+k+1)U^a_k=-\sum\limits_{l=1}^{k-2}\left[\mc{M}_{k,l}U^a_l\right.\\
&\left.+\sum\limits_{r,s=1}^{N-1}\left(\frac{1}{2}a(a+1)+s(s+1)\right)d^a_{rs}U^r_lU^s_{k-l-1}\right.\\
&\left.+\frac{1}{2}\sum\limits_{p=2}^{k-l-2}\sum\limits_{r,s,t,u=1}^{N-1}\left(u(u+1)d^a_{ru}d^u_{rs}U^r_lU^s_{p-1}U^t_{k-l-p-1}\right)\right.\\
&\left.-\hat{M}_{l+1}\left[(a-k+l-1)(a+k-l)U^a_{k-l-1}\right.\right.\\
&\left.\left.+\slim_{p=2}^{k-l-2}\left(\slim_{r,s=1}^{N-1}\left(\frac{1}{2}a(a+1)+s(s+1)\right)d^a_{rs}U^r_{p-1}U^s_{k-l-p-1}\right.\right.\right.\\
&\left.\left.\left.+\frac{1}{2}\slim_{r=2}^{k-l-p-2}\slim_{s,t,u,v=1}^{N-1}\left(v(v+1)d^a_{sv}d^v_{ut}U^s_{p-1}U^t_{r-1}U^u_{k-l-p-r-1}\right)\right)\right]\right.\\
&\left.+\frac{\lambda_N^2}{4S_0^2}\slim_{r=1}^{N-1}\left(E^r_lE^a_{k-l-1}v^i_r+\slim_{s=1}^{N-1}\slim_{p=2}^{k-l-2}U^a_{p-1}E^r_lE^s_{k-l-p-1}v^i_rv^i_s\right)\right].
\end{split}
\end{equation}

Now, for the lower order terms we note from \eqref{Ukk} that
\begin{equation}
E^1_1=\theta_1, \quad E^2_2=\theta_2, \quad U^1_1=\beta_1, \quad U^2_2=\beta_2,
\end{equation}
and from the right hand sides of (\ref{Eak}, \ref{Uak}) we see that
\begin{equation}
E^a_1=U^a_1=0\mbox{ for }a>1, \quad E^a_2=U^a_2=0\mbox{ for }a>2;\\
\end{equation}
so that the first two terms in each gauge field expansion are defined by one parameter each. What we wish to prove is that this is true in general, and that every new term in the expansion is determined by one new parameter each, until $N-1$ parameters have been introduced for each gauge field (note that this is not assumed -- it emerges naturally from the size of matrix \eqref{matA}). Therefore we prove the following:

\begin{prop}\label{UEprop}
$E^a_k=U^a_k=0$ for $a>k$.
\end{prop}
\textbf{Proof} In order to show this, we use a form of 2-proposition `ladder' induction, defined by the following sentences:
\begin{equation}
\begin{split}
E(k):\mbox{ ``}E^a_\nu=0,\mbox{ for }a>\nu,\mbox{ for \underline{each} }\nu=2,...,k."\\
U(k):\mbox{ ``}U^a_\nu=0,\mbox{ for }a>\nu,\mbox{ for \underline{each} }\nu=2,...,k."
\end{split}
\end{equation}
Our strategy is as follows.
\begin{itemize}
\item We know that $E(2)\land U(2)$ is true.
\item We make the induction hypothesis $E(\kappa)\land U(\kappa)$ for some $\kappa\in\nat$, $2<\kappa<N-1$. 
\item We show that $E(\kappa)\land U(\kappa)\implies E(\kappa+1)$.
\item We show that $E(\kappa+1)\land U(\kappa)\implies U(\kappa+1)$.
\item Thus we have shown that $E(\kappa)\land U(\kappa)\implies E(\kappa+1)\land U(\kappa+1)$ and the induction is complete: $E(k)\land U(k)$ is true for all $k\geq 2$.
\end{itemize}
We will need to make use of lemma 4 in K\"{u}nzle \cite{HPK1994_analysis_sun}, that is,
\begin{equation}\label{dars}
d^a_{rs}=0\mbox{ if }r+s\leq a,
\end{equation}
in order to endure that strategic terms of \eqref{Eak} and \eqref{Uak} vanish. In addition to this fact we will need another analogous fact:
\begin{equation}\label{darst}
d^a_{rst}\equiv d_a^{-1}(N-1)^3c_{arst}=0\mbox{ if }r+s+t\leq a
\end{equation}
where $c_{ijkl}$ is defined below. We shall now briefly prove this fact, by way of a lemma.

\begin{lem}
\begin{equation}
c_{ijkl}\equiv\slim_{r=1}^{N-1}\frac{r}{(N-r)^3}Q_i(r-1)Q_j(r-1)Q_k(r-1)Q_l(r-1)=0,
\end{equation}
if $i+j+k\leq l$.
\end{lem}
\vsp
\textbf{Proof} First we note that $c_{ijkl}$ is symmetric on all of its indices. We shall use the following induction argument:
\begin{itemize}
\item We show that $c_{11kl}=0$ for $2+k\leq l$;
\item We make the induction hypothesis that $1+j+k\leq l$ implies $c_{1jkl}=0$, for all $j\in\{1,...,J\}$ for some $J\in\nat$, $1<J<N-1$;
\item We show that $c_{1Jkl}=0$ for $1+J+k\leq l$ implies that $c_{1,J+1,kl}=0$ for $2+J+k\leq l$;
\item Finally we note that because $c_{ijkl}$ is symmetric in its first two indices, we are done; since the line above implies that if $c_{I,1,kl}=0$ for $I+1+k\leq l$, then $c_{I+1,1,kl}=0$ for $I+2+k\leq l$, and therefore induction over one index is induction over both indices. Hence we conclude that 
\begin{equation}
c_{ijkl}=0\mbox{ if }i+j+k\leq l.
\end{equation}
\end{itemize}
Firstly, note \cite{SK1961_Hahn} that $Q_1(r-1)=\frac{N-r}{N-1}$, so that
\begin{equation}
\begin{split}
c_{11kl}=&\slim_{r=1}^{N-1}\frac{r}{(N-r)(N-1)^2}Q_k(r-1)Q_l(r-1)\\
=&\frac{1}{(N-1)^3}d_k\kron^k_l=0\mbox{ if }k\neq l,\\
\end{split}
\end{equation}
e.g. if $2+k\leq l$.

Next, assume that
\begin{equation}\label{c1jkl}
c_{1jkl}=\slim_{r=1}^{N-1}\frac{r}{(N-r)^2(N-1)}Q_j(r-1)Q_k(r-1)Q_l(r-1)=0
\end{equation}
if $1+j+k\leq l$, for all $j\in\{1,...,J\}$, $1<J<N-1$. Now we use the recurrence relation for the Hahn polynomials
\begin{equation}
-rQ_i(r)=d_iQ_{i-1}(r)+(b_i+d_i)Q_i(r)+b_iQ_{i+1}(r)
\end{equation}
where
\begin{equation}
b_i=\frac{(i+1)(N-i-1)}{2(2i+1)}, \quad d_i=\frac{(i-1)(N+i)}{2(2i+1)},
\end{equation}
in \eqref{c1jkl}: first with $Q_i(r-1)$, then with $Q_k(r-1)$, then we let $j\mapsto j-1$ and equate the expressions. This yields
\begin{equation}\label{crr}
\begin{split}
c_{1jkl}=&\frac{1}{b_{j-1}}\left[-d_{j-1}c_{1,j-2,kl}+d_lc_{1,j-1,kl}\right.\\
&\left.+(b_{j-1}+d_{j-1}-b_l-d_l)c_{1,j-1,kl}+b_lc_{1,j-1,k,l+1}\right].\\
\end{split}
\end{equation}
Now to complete the induction step, we let $j=J+1$, with the restriction $2+J+k\leq l$, i.e. $J+k\leq l-2$, and consider each term of \eqref{crr}:
\begin{itemize}
\item $c_{1,J-1,kl}=0$, since $1+(J+1)+k\leq l$, i.e. $J+k\leq l-2$;
\item $c_{1,J,k,l-1}=0$, since $1+J+k\leq l$, i.e. $J+k\leq l-1$, and we know $J+k\leq l-2\leq l-1$;
\item $c_{1,J,k,l}=0$, since $1+J+k\leq l$, i.e. $J+k\leq l-1$, and  we know $J+k\leq l-2\leq l-1$;
\item $c_{1,J,k,l+1}=0$, since $1+J+k\leq l+1$, i.e. $J+k\leq l$, and we know $J+k\leq l-2\leq l$.
\end{itemize}
Therefore, all terms on the right hand side of \eqref{crr} vanish and hence $c_{1jkl}=0\,\,\forall_{j,k,l}\in\{1,...,N-1\}$ if $1+j+k\leq l$. Finally, using the symmetry argument described earlier, we can say that $c_{i1kl}=0\quad\forall_{i,k,l}\in\{1,...,N-1\}$ if $i+1+k\leq l$, and hence the lemma is proven.$\Box$

\vsp
\begin{cor} This means, using \eqref{evec}, that we can write
\begin{equation}
d^a_{rst}=d_a^{-1}(N-1)^3c_{arst}=0\mbox{ if }r+s+t\leq a.
\end{equation}
Thus we have proven \eqref{darst} true.
\end{cor}

So now we are equipped to take \eqref{Eak} and \eqref{Uak} in turn, making the appropriate induction hypothesis $E(\kappa)\land U(\kappa)$ for some $2<\kappa<N-1$; hence we now substitute $k=\kappa+1$ into each term of \eqref{Eak} and prove that $E(\kappa)\land U(\kappa)\implies E(\kappa+1)$.
\begin{itemize}
\item $\slim_{l=1}^{\kappa-1}\mc{N}_{\kappa+1,l}E^a_l=0$ since $l\leq\kappa-1<\kappa+1$.
\item $2a(a+1)\slim_{l=1}^{\kappa-1}\slim_{r=1}^l\slim_{c=1}^{\kappa-l}d^a_{rs}U^r_lE^s_{\kappa-l}$:\\
Now $r+s\leq\kappa<\kappa+1$, so that $d^a_{rs}=0$ here.
\item $a(a+1)\slim_{l=1}^{\kappa-1}\slim_{p=1}^{\kappa-l-1}\slim_{r=1}^{p-1}\slim_{s=1}^l\slim_{t=1}^{\kappa-l-p}d^a_{rst}U^r_{p-1}U^s_lE^t_{\kappa-l-p}$:\\
Here, $r+s+t\leq \kappa-1<\kappa+1$ so $d^a_{rst}=0$.
\item $\slim_{l=1}^{\kappa-1}S_{l+1}(a+k-l)(a-k+l+1)E^a_{\kappa-l}=0$ since $\kappa-l\leq \kappa-1$.
\item $2a(a+1)\slim_{l=1}^{\kappa-1}S_{l+1}\slim_{p=2}^{\kappa-l-1}\slim_{r=1}^{p-1}\slim_{s=1}^{\kappa-l-p}d^a_{rs}U^r_{p-1}E^s_{\kappa-l-p}$.\\
This is similar to term 2: we get $r+s\leq \kappa-l-1\leq \kappa-2$, hence $d^a_{rs}=0$ here.
\item $a(a+1)\slim_{l=1}^{\kappa-1}S_{l+1}\slim_{p=2}^{\kappa-l-1}\slim_{r=2}^{\kappa-l-p-1}\slim_{q=1}^{r-1}\slim_{s=1}^{p-1}\slim_{t=1}^{\kappa-l-p-r}d^a_{qst}U^q_{r-1}U^s_{p-1}E^t_{\kappa-l-p-r}$.\\
Now $q+s+t\leq \kappa-l-2\leq \kappa-3$. Therefore $d^a_{qst}=0$ here too.
\end{itemize}
Hence we can say that $E(\kappa)\land U(\kappa)\implies E(\kappa+1)$. Now we must take \eqref{Uak} to prove the other statement: that $E(\kappa+1)\land U(\kappa)\implies U(\kappa+1)$.

It is clear from \eqref{Uak} that the left hand side and the first two terms on the right hand side are identical to those previously examined \cite{HPK1994_analysis_sun}, so these are already proven to be zero under this process, especially as that examination \cite{HPK1994_analysis_sun} makes no mention on the form of $\mc{M}_{j,k}$ and so the appearance of $\ell$ is unimportant \cite{BW2008_sun_ex}. The next three terms are very similar, but we'll examine them anyway.
\begin{itemize}
\item $\slim_{l=1}^{\kappa-1}\hat{M}_{l+1}(a-k+l+1)(a+k-l)U^a_{\kappa-l}=0$ since $\kappa-l\leq\kappa-1$.
\item $\slim_{l=1}^{\kappa-1}\hat{M}_{l+1}\slim_{p=2}^{\kappa-l-1}\slim_{r=1}^{p-1}\slim_{s=1}^{\kappa-l-p}\left(\frac{1}{2}a(a+1)+s(s+1)\right)d^a_{rs}U^r_pU^s_{\kappa-l-p}$:\\
Now $r+s\leq \kappa-l-1\leq \kappa-2$, so $d^a_{rs}=0$ here.
\item $\frac{1}{2}\slim_{l=1}^{\kappa-1}\hat{M}_{l+1}\slim_{p=2}^{\kappa-l-1}\slim_{r=2}^{\kappa-l-p-1}\slim_{s=1}^{p-1}\slim_{t=1}^{r-1}\slim_{u=1}^{\kappa-l-p-r}\slim_{v=1}^{N-1}v(v+1)d^a_{sv}d^v_{ut}U^s_{p-1}U^t_{r-1}U^u_{\kappa-l-p-r}$:\\
Here $u+t\leq \kappa-l-p-1$, therefore $d^v_{ut}=0$ unless $v<\kappa-l-p-1$. But then, $v+s\leq \kappa-l-2\leq\kappa-3$. So $d^a_{sv}=0$.
\end{itemize}
The final two terms are less familiar but present little problem.
\begin{itemize}
\item $\disfrac{\lambda_N^2}{4S_0^2}\sum\limits_{l=1}^{\kappa-1}\slim_{r=1}^{N-1}E^r_lE^a_{\kappa-l}v^i_r$:\\
In this term, note that we have $E^a_{\kappa-l}$ in the product. But $l\geq 1$ and so $\kappa-l\leq\kappa-1$ and this term is zero.
\item $\disfrac{\lambda_N^2}{4S_0^2}\sum\limits_{l=1}^{\kappa-1}\slim_{r,s=1}^{N-1}\slim_{p=2}^{\kappa-l-1}U^a_{p-1}E^r_lE^s_{\kappa-l-p}v^i_rv^i_s$:\\
Finally, we note the $U^a_{p-1}$ multiplicand in this term. However we also have $p-1\leq\kappa-l-2\leq\kappa-3$, so once again this term is zero.
\end{itemize} 

Hence we have shown that $E(\kappa+1)\land U(\kappa)\implies U(\kappa+1)$. Thus the induction is complete, and Proposition \ref{UEprop} is proven.$\Box$

\vsp
Therefore, we can now finally express $u_j$ and $\mc{E}_j$ as finite power series, in vector form:
\begin{equation}\label{ueexp}
\begin{split}
\bsym{u}(x)=&\bsym{u}_0+\slim_{k=1}^{N-1}\bsym{\beta}_k(x)x^{k+1},\\
\bsym{\mc{E}}(x)=&\slim_{k=1}^{N-1}\bsym{\theta}_k(x)x^k,\\
\end{split}
\end{equation}
where $\bsym{u}_0=(1,1,...,1)^T$ is an $N-1$-vector and
\begin{equation}\label{betathetaUE}
\begin{array}{l}
\bsym{\beta}_k(x)\equiv\bar{\beta}_k(x)\bsym{v}_k\equiv\slim_{l=k}^{N-1}U^k_l\bsym{v}_kx^{l-k},\\
\bsym{\theta}_k(x)\equiv\bar{\theta}_k(x)\bsym{v}_k\equiv\slim_{l=k}^{N-1}E^k_l\bsym{v}_kx^{l-k}.\\
\end{array}
\end{equation}
We define $2(N-1)$ initial parameters $\{\breve{\beta}_k,\breve{\theta}_k\}$ where 
\begin{equation}\label{breve}
\breve{\beta}_k\equiv\bar{\beta}_k(0), \quad \breve{\theta}_k\equiv\bar{\theta}_k(0),
\end{equation}
in which the constants $U^k_l$ and $E^k_l$ depend quite complicatedly on the constants $\{\breve{\beta}_k,\breve{\theta}_k\}$. To obtain the boundary conditions \eqref{origexp} then, we substitute $u_j$ back into the expression for $\omega_j$, and augment them both with possible higher order terms. Therefore we have shown that the gauge field equations are entirely specified by $2N-2$ initial parameters which can be chosen in a consistent way.

Finally, we finish off our proof of Proposition \ref{conprop} with the following argument. Assume that for some $k$ we know all the parameters $m_3,...,m_{k-1}$, $S_0,...,S_{k-1}$, $u_{j,2},...,u_{j,k-1}$ and $\mc{E}_{j,1},...,\mc{E}_{j,k-1}$. Using the recurrence relation \eqref{recm} we can certainly work out $m_k$. The detailed study of the Yang-Mills equations in the form \eqref{YMsys} has shown that we can work out $u_{j,k}$ and $\mc{E}_{j,k}$. Finally, we can use the final recurrence relation \eqref{recS} to work out $S_k$, and Proposition \ref{conprop} is proven.$\Box$

\vsp
Before we go on to prove local existence near the origin, we note an interesting consequence of \eqref{betathetaUE}. It may be noted that the dependence of both $\bar{\beta}_k(x)$ and $\bar{\theta}_k(x)$ on $x$ is the factor $x^{k-l}$, hence we might expect that in fact $\bar{\beta}_k(x)\propto\bar{\theta}_k(x)$ for each $k$. This is fairly easy to see once it is realised that we have expanded both gauge fields in the same basis of eigenvectors, but to be precise, we can prove a minor lemma that will help confirm our results later.

\begin{lem}\label{betaproptheta}
\begin{equation}
\bsym{u}_{k+1}=\tau_k\bsym{\mc{E}}_k\mbox{ for $N-1$ constants }\tau_k,
\end{equation}
for $k\in\{1,...,N-1\}$, and hence
\begin{equation}
\bsym{\beta}_k=\tau_k\bsym{\theta}_k,
\end{equation}
no sum over $k$.
\end{lem}
\vsp
\textbf{Proof} Take the second equation in each of \eqref{umat} and \eqref{emat}:
\begin{equation}
\bsym{\sigma}^k\bsym{u}_{k+1}=d_k\beta_k, \quad \bsym{\sigma}^k\bsym{\mc{E}}_k=d_k\theta_k,
\end{equation}
(for $d_k$ defined in \eqref{dk}) and define $\tau_k\equiv\frac{\beta_k}{\theta_k}$ (no sum on $k$). Then it is clear that
\begin{equation}
\bsym{\sigma}^k\bsym{u}_{k+1}=\tau_k\bsym{\sigma}^k\bsym{\mc{E}}_k.
\end{equation}
Since this is a scalar product and we have not mentioned the form of $\bsym{\sigma}^k$ at all, this must hold for each pair of vectors $\bsym{u}_{k+1}$ and $\bsym{\mc{E}}_k$, i.e.
\begin{equation}
\bsym{u}_{k+1}=\tau_k\bsym{\mc{E}}_k,
\end{equation}
and immediately, 
\begin{equation}
\bsym{\beta}_k=\tau_k\bsym{\theta}_k,\mbox{ with }\tau_k=\frac{\beta_k}{\theta_k}
\end{equation}
(no sum on $k$).$\Box$

\vsp
Notice this is \textit{not} saying that the functions $u_j$ and $\mc{E}_j$ are multiples of each other; rather that at each order, the expansion vectors will be at most a scalar multiple different from each other. Later, this gives us a form to aim for when rearrange the equations to make use of Theorem \ref{exdetheor}.

\subsection{Local existence of solutions at the origin $r=0$ ($k=1$ only)}

Now we can use the results obtained in Section \ref{lexboundsec} to prove the existence of solutions regular at the origin and analytic in their initial parameters. First we must rewrite the equations in a form applicable to Theorem \ref{exdetheor}. Hence we will state our proposition:

\begin{prop}\label{lex0}
There exists a $2N-2$-parameter family of solutions to the field equations in the form (\ref{EE1}, \ref{EE2}, \ref{YM1E}, \ref{YM2E2}) which are regular at $r=0$. These solutions are analytic in some neighbourhood of $r=0$ and in their initial conditions $\{\breve{\beta}_k,\breve{\theta}_k\}$ \eqref{breve}.
\end{prop}
\vsp
\textbf{Proof} We begin with the scaled equations from Subsection \ref{psr0con} \eqref{FEr02} (\textit{sans} tildes), except that we take the original definition of $S$ so that the first term of the expansion is $S_0$. 

It is best to begin by examining the equation for the MGFs. Introducing the vectors $\bsym{u}\equiv(u_1,u_2,...,u_{N-1})^T$, $\boldsymbol{\mc{E}}\equiv(\mc{E}_1,\mc{E}_2,...,\mc{E}_{N-1})^T$, we may write this equation as
\begin{equation}
x^2\mu\frac{d^2\bsym{u}}{dx^2}+2\left(m-x\kappa_NP+\frac{x^3}{\ell^2}-\frac{x^3\eta}{4S^2}\right)\frac{d\bsym{u}}{dx}+\frac{1}{2}\bsym{\mc{W}}+\frac{\lambda_N^2x^2}{4\mu S^2}\bsym{\mc{V}}=0;
\end{equation}
introducing two $N-1$ vectors
\begin{equation}
\begin{split}
\bsym{\mc{W}}\equiv&(\mc{W}_1,\mc{W}_2,...,\mc{W}_{N-1})^T\mbox{ with }\mc{W}_j=\left(q_{j+1}-q_j\right)u_j,\\
\bsym{\mc{V}}\equiv&(\mc{V}_1,\mc{V}_2,...,\mc{V}_{N-1})^T\mbox{ with }\mc{V}_j=u_j\mc{E}_j^2.
\end{split}
\end{equation}
We notice that this is very similar to the corresponding purely magnetic form \cite{HPK1994_analysis_sun}, so parts of the analysis will carry over similarly. Using \eqref{matA}, we can write $\mc{W}_j$ as
\begin{equation}
\mc{W}_j=2u_j-\slim_{i=1}^{N-1}u_jA^i_ju_i^2\quad\mbox{ (No sum on } j).
\end{equation}
Also, we make use of the fact that as functions defined with the eigenvectors of $A^i_j$, $\bsym{\beta}_k(x)$ and $\bsym{\theta}_k(x)$, will satisfy
\begin{equation}\label{Abethe}
A\bsym{\beta}_k(x)=k(k+1)\bsym{\beta}_k(x), \quad A\bsym{\theta}_k(x)=k(k+1)\bsym{\theta}_k(x).
\end{equation}
We should state that we begin by assuming nothing about the properties of the functions $\bsym{\beta}_k(x)$ or $\bsym{\theta}_k(x)$ at this stage; for instance we do not yet assume they are regular at $x=0$ nor that they are of the form \eqref{betathetaUE}.

Multiplying by the left eigenvectors $\bsym{\sigma}^k$ (which we assume to be constant and independent of $x$) gives:
\begin{equation}\label{YM2sig}
x^2\mu\frac{d^2}{dx^2}(\bsym{\sigma}^k\bsym{u})+2\left(m-x\kappa_NP+\frac{x^3}{\ell^2}-\frac{x^3\eta}{4S^2}\right)\frac{d}{dx}(\bsym{\sigma}^k\bsym{u})+\frac{1}{2}\bsym{\sigma}^k\bsym{\mc{W}}+\frac{\lambda_N^2x^2}{4\mu S^2}\bsym{\sigma}^k\bsym{\mc{V}}=0.
\end{equation}
Using the orthogonality of eigenvectors, whence $\bsym{\sigma}^A\bsym{\beta}_B(x)=0$ iff $A\neq B$, we define $N-1$ functions $\bsym{\sigma}^k\bsym{\beta}_k(x)\equiv\xi^1_k(x)$ (no sum on $k$). Using \eqref{ueexp}, this transforms \eqref{YM2sig} into
\begin{equation}\label{YM2sig2}
\begin{split}
0=&x^2\mu\left(x^{k+1}\frac{d^2\xi^1_k}{dx}+2(k+1)x^k\frac{d\xi^1_k}{dx}+ k(k+1)x^{k-1}\xi^1_k(x)\right)\\
&+2\left(m-\kappa_NxP+\frac{x^3}{\ell^2}-\frac{x^3\eta}{4S^2}\right)\left(x^{k+1}\frac{d\xi^1_k}{dx}+(k+1)x^k\xi^1_k(x)\right)\\
&+\frac{1}{2}\bsym{\sigma}^k\bsym{\mc{W}}+\frac{\lambda_N^2x^2}{\mu S^2}\bsym{\sigma}^k\bsym{\mc{V}}.
\end{split}
\end{equation}
The final term involving $\bsym{\mc{V}}$ does not cause concern since it is of quite high order; the complicated part of the analysis is resetting $\bsym{\mc{W}}$ in a suitable form. Fortunately we may appeal to two lemmata courtesy of K\"{u}nzle and Oliynyk \cite{OK2002_gen_spher_EYM} (see equations (7.3) and just below (7.7)):
\begin{lem}\label{Wlem}
\begin{equation}\label{Wleme}
\frac{1}{2}\bsym{\mc{W}}=-\slim_{k=1}^{N-1}k(k+1)\bsym{\beta}_k(x)x^{k+1}+\slim_{l=2}^z\bsym{g}_lx^l
\end{equation}
for vectors $\bsym{g}_l$ and some integer $z>2$. $\Box$
\end{lem}

The details here are unimportant (we shall show an analogous proof later), as are the precise value of $z$ or the forms of $\bsym{g}_l$: the key point in this result is that $\bsym{\mc{W}}\sim O(x^2)$ at least. In addition, we have

\begin{lem}\label{Wlem2}
\begin{equation}\label{Wlem2e}
\bsym{\sigma}^k\bsym{g}_{l}=0\mbox{ if }l<k+2.\mbox{ }\Box
\end{equation}
\end{lem}
Concerning the final term in \eqref{YM2sig2}, we may establish that
\begin{equation}
\bsym{\sigma}^k\bsym{\mc{V}}=x^{k+3}\xi^1_k\slim_{l=1}^{N-1}\slim_{p=1}^{N-1}\bsym{\theta}_l\bsym{\theta}_px^{l+p-2}.
\end{equation}
If we introduce new variables
\begin{equation}\label{dxi1}
\frac{d\xi^1_k}{dx}\equiv\psi^1_k(x),
\end{equation}
divide \eqref{YM2sig2} through by $x^{k+2}\mu$, rearrange and use (\ref{Wleme}, \ref{Wlem2e}) we can finally rewrite \eqref{YM2sig2} as
\begin{equation}\label{dpsi1}
\begin{split}
x\frac{d\psi^1_k}{dx}=&-2(k+1)\psi^1_k-k(k+1)\xi^1_kx^{-1}+k(k+1)\xi^1_k\frac{x^{-1}}{\mu}\\
&-\frac{2}{x^2\mu}\left(m-\kappa_NxP+\frac{x^3}{\ell^2}-\frac{x^3\eta}{4S^2}\right)\left((k+1)\xi^1_k+x\psi^1_k\right)\\
&-\frac{\lambda_N^2x^3}{4\mu^2S^2}\xi^1_k\slim_{l=1}^{N-1}\slim_{p=1}^{N-1}\bsym{\theta}_l\bsym{\theta}_px^{l+p-2}-\mu^{-1}\slim_{l=0}^{z-k-2}\bsym{\sigma}^k\bsym{g}_{l+k+2}x^l.\\
\end{split}
\end{equation}

Now we leave the Yang-Mills equations for a while as we must consider the expansions of the various quantities $G$, $q_j$, $P$, $\eta$ and $\zeta$. In order to do this we must note that the first eigenvector $\bsym{v}_1=(1,1,...,1)^T$. This is easy to prove if we observe that
\begin{equation}
v_1^i=\frac{N-1}{N-i}\,_2F_1(-1,-i+1;-N+1;1),
\end{equation}
and then use the special case of Gauss' theorem; that is,
\begin{equation}
_2F_1(-n,b;c;1)=\frac{(c-b)_{(n)}}{(c)_{(n)}}\mbox{ for }\mk{Re}(c)\leq\mk{Re}(-n+b),
\end{equation}
for $n\in\nat^+$ and where $(a)_{(n)}$ is the Pochhammer symbol defined by $(a)_{(0)}=1$, $(a)_{(n)}=\prod\limits_{i=0}^{n-1}(a+i)\,\,(n>0)$ \cite{SK1961_Hahn}. This means that all $u_{j,2}$ are equal for all $j$, and similar for $\mc{E}_{j,1}$. Hence we write
\begin{equation}
\begin{array}{lcl}
u_{j,2}=\beta_1(x), & \quad & \mc{E}_{j,1}=\theta_1(x),
\end{array}
\end{equation}
(for all $j$) and we introduce the expansions
\begin{equation}
\beta_1(x)\equiv\beta_0+x\hat{\beta}_1(x), \quad \theta_1(x)\equiv\theta_0+x\hat{\theta}_1(x),
\end{equation}
where again, we assume nothing about the functions $\hat{\beta}_1(x)$ or $\hat{\theta}_1(x)$; they can be written in terms of $\xi^1_k$ and $\xi^2_k$ but the precise expressions are not important. Note also that from now on, a hat $\hat{ }$ indicates a function we have introduced to deal with higher order behaviour, which is analytic and regular in all transformed field variables with which they are defined. Thus we obtain
\begin{equation}\label{quanexp}
\begin{split}
G&=\frac{4\beta_0^2}{\kappa_N}x^2+x^3\hat{G},\\
q_j&=2(N-1+2j)x^2\beta_0+x^3\hat{q_j},\\
P&=\frac{2\beta_0^2}{\kappa_N}x^2+x^3\hat{P},\\
\zeta&=\frac{\theta_0^2}{\kappa_N}x^2+x^3\hat{\zeta},\\
\eta&=\frac{\theta_0^2}{2\kappa_N}+x\hat{\eta}.\\
\end{split}
\end{equation}
Now we can finally tackle the Einstein equations in \eqref{FEr03}. Using \eqref{quanexp}, they may be written
\begin{equation}\label{xds}
\begin{split}
x\frac{dm}{dx}&=3\left(2\beta_0^2+\frac{\theta_0^2}{8\kappa_N}\right)x^3+x^4\hat{m},\\
x\frac{dS}{dx}&\equiv x^2\hat{S}_A.
\end{split}
\end{equation}
The equation for $S$ is now in the required form, and we note by the theorem that this implies the expansion is of the form $S(x)=S_0+O(x^2)$, a fact we will need momentarily. 

If we define a new variable for $m$ by
\begin{equation}
\chi=\frac{1}{x^3}\left(m-m_3x^3\right)\mbox{ with }m_3\equiv2\beta_0^2+\frac{\theta_0^2}{8\kappa_N},
\end{equation}
then we find
\begin{equation}\label{xdchi}
x\frac{d\chi}{dx}=-3\chi+x\hat{m}.
\end{equation}
Note also that $\mu=1+x^2(2m_3-2\chi+\frac{1}{\ell^2})$. This equation is now also in our desired form, and from Theorem \ref{exdetheor} we get the behaviour $\chi=O(x)$ and hence $m=m_3x^3+O(x^4)$.

Now we can go back to the MGF equation \eqref{dpsi1}. Using \eqref{xds}, and writing $\mu^{-1}=1+x^2\hat{\mu}$, $S^{-1}=S_0^{-1}+x^2\hat{S}_B$, and $S^{-2}=S_0^{-2}+x^2\hat{S}_C$, we get
\begin{equation}
\begin{split}
-2\left(m-x\kappa_NP+\frac{x^3}{\ell^2}-\frac{x^3\eta}{4S^2}\right)&\equiv x^3\hat{\mc{H}}_0,\,\,\mbox{ and}\\
-\frac{\lambda_N^2x^3}{4\mu^2S^2}\xi^1_k\slim_{l=1}^{N-1}\slim_{p=1}^{N-1}\bsym{\theta}_l\bsym{\theta}_px^{l+p-2}&\equiv x^3\hat{\mc{H}}_{1,k}.\\
\end{split}
\end{equation}
Therefore we can write \eqref{dpsi1} as
\begin{equation}
x\frac{d\psi^1_k}{dx}=-2(k+1)\psi^1_k+x\hat{\mc{H}}_{2,k}-\bsym{\sigma}^k\bsym{g}_{k+2}.
\end{equation}
We make one final change of variables here, as we let
\begin{equation}\label{barpsi1}
\bar{\psi}^1_k\equiv\psi^1_k+\frac{1}{2(k+1)}\bsym{\sigma}^k\bsym{g}_{k+2},
\end{equation}
and finally get our equations in the desired form:
\begin{equation}\label{xdpsik}
x\frac{d\bar{\psi}^1_k}{dx}=-2(k+1)\bar{\psi}^1_k+x\hat{\mc{H}}_{2,k}.
\end{equation}

Now we turn our attention to the equation for $\mc{E}_j$ in \eqref{FEr02}. We shall need to define analogous transforms to those for $\xi^1_k$ and $\psi^1_k$, that is
\begin{align}
\bsym{\sigma}^k\bsym{\theta}_k(x)&\equiv\xi^2_k(x)\,\,\mbox{ (no sum on $k$)},\label{xi2}\\
\frac{d\xi^2_k}{dx}&\equiv\psi^2_k(x).\label{dxi2}
\end{align}
We recall that lemma \ref{betaproptheta} established that $\bsym{\beta}_k=\tau_k\bsym{\theta}_k$ for some constants $\tau_k$. Therefore, we can deduce that $\xi^1_k=\tau_k\xi^2_k$ and $\psi^1_k=\tau_k\psi^2_k$; and hence if $\xi^1_k$ and $\frac{d\xi^1_k}{dx}$ are regular and analytic near $x=0$, then so should $\xi^2_k$ and $\frac{d\xi^2_k}{dx}$ be, and they should exhibit the same behaviour at the origin as $\xi^1_k$ and $\frac{d\xi^1_k}{dx}$, as dictated by Theorem \ref{exdetheor}. We do not feel that completes the proof; rather it gives us a method of confirming our result for the analogue of \eqref{xdpsik} for the EGFs, since it means the equations for the two gauge fields should look almost identical.

We begin with the EGF equations in the following vector form:
\begin{equation}
x^2\mu\frac{d^2\bsym{\mc{E}}}{dx^2}=x\mu\left(\frac{x}{S}\frac{dS}{dx}-2\right)\frac{d\bsym{\mc{E}}}{dx}+\bsym{\mc{Z}},
\end{equation} 
with $\bsym{\mc{Z}}$ in component form as
\begin{equation}
\mc{Z}_j=2\gamma_ju_j^2\mc{E}_j-\gamma_{j-1}u_{j-1}^2\mc{E}_{j-1}-\gamma_{j+1}u_{j+1}^2\mc{E}_{j+1}.
\end{equation}
We proceed as before: we multiply through by the left eigenvectors $\bsym{\sigma}_k$ and substitute in our new variable $\xi^2_k$ \eqref{xi2}, which gives us
\begin{equation}\label{xi2e}
\begin{split}
0=&x^2\mu\left(x^k\frac{d^2\xi^2_k}{dx^2}+2kx^{k-1}\frac{d\xi^2_k}{dx}+k(k-1)x^{k-2}\xi^2_k(x)\right)\\
&-x\mu\left(\frac{x}{S}\frac{dS}{dx}-2\right)\left(x^k\frac{d\xi^2_k}{dx}+kx^{k-1}\xi^2_k(x)\right)-\bsym{\sigma}^k\bsym{\mc{Z}}.\\
\end{split}
\end{equation}
Similar to before, the troublesome part is writing $\bsym{\mc{Z}}$ in a form that renders it applicable to Theorem \ref{exdetheor}; but similarly, it may be expressed in terms of the matrix $A^i_j$ \eqref{matA}, and thus we may formulate two lemmata analogous to \ref{Wlem} and \ref{Wlem2}, found in K\"{u}nzle and Oliynyk \cite{OK2002_gen_spher_EYM}:

\begin{lem}
\begin{equation}\label{Zlem}
\bsym{\mc{Z}}=\slim_{k=1}^{N-1}k(k+1)\bsym{\theta}_kx^k+\slim_{l=3}^y\bsym{f}_lx^l
\end{equation}
for some constant $y\in\nat^+,y>3$.
\end{lem}
\vsp
\textbf{Proof} We may write $\bsym{\mc{Z}}$ in components as
\begin{equation}
\mc{Z}_j=\slim_{i=1}^{N-1}A^i_ju^2_i\mc{E}_i.
\end{equation}
We let $\tilde{u}_i=u_i-1$, so that the lowest order term in $\tilde{u}_i$ is order $x^2$. Thus
\begin{equation}\label{Zinutilde}
\mc{Z}_j=\slim_{i=1}^{N-1}A^i_j(\mc{E}_i+2\tilde{u}_i\mc{E}_i+\tilde{u}^2_i\mc{E}_i).
\end{equation}
Consider
\begin{equation}
\slim_{i=1}^{N-1}A^i_j\mc{E}_i=\slim_{i=1}^{N-1}\slim_{k=1}^{N-1}A^i_j\bsym{\theta}_kx^k.
\end{equation}
Using the second equation in \eqref{Abethe}, we obtain
\begin{equation}
\slim_{i=1}^{N-1}A^i_j\mc{E}_i=\slim_{k=1}^{N-1}k(k+1)\bsym{\theta}_kx^k.
\end{equation}
Finally we note that the other terms in \eqref{Zinutilde} are at least of order $x^3$, and therefore our result follows.$\Box$

\vsp
Once again the value of $y$ and the form of the vectors $\bsym{f}_l$ are unimportant compared to the form of \eqref{Zlem}. We also prove the following:
\begin{lem}
\begin{equation}\label{Zleme2}
\bsym{\sigma}^k\bsym{f}_l=0\mbox{ if } l<k+2.
\end{equation}
\end{lem}
\vsp
\textbf{Proof} Following the previous lemma we may consider $\bsym{\mc{Z}}$ \eqref{Zinutilde} as a sum of three parts:
\begin{equation}
\begin{split}
\mc{Z}_j=&\slim_{i=1}^{N-1}A^i_j\mc{E}_i+2\slim_{i=1}^{N-1}A^i_j\tilde{u}_i\mc{E}_i+\slim_{i=1}^{N-1}A^i_j\tilde{u}^2_i\mc{E}_i.
\end{split}
\end{equation}
We use the previous eigenvector expansions of the matrix $A^i_j$ \eqref{matA}, the definitions (\ref{vwnorm}, \ref{dars}, \ref{darst}) and the second equation in \eqref{Abethe} to give (in components)
\begin{equation}
\begin{split}
\sigma^k_j\mc{Z}_j=&k(k+1)\xi^2_k(x)x^k\\
&+2k(k+1)\slim_{l=2}^{2N-1}\slim_{p=1}^{l-2}d_kd^k_{p,l-p-1}\bar{\theta}_p(x)\bar{\beta}_{l-p-1}(x)x^l\\
&+k(k+1)\slim_{l=3}^{3N-1}\slim_{p=1}^{l-4}\slim_{q=1}^{l-p-2}d_kd^k_{q,p,l-p-q-2}\bar{\theta}_q(x)\bar{\beta}_p(x)\bar{\beta}_{l-p-q-2}(x)x^l\\
\end{split}
\end{equation}
(where we note that $\bar{\beta}_k(x)$ and $\bar{\theta}_k(x)$ are zero for $k\geq N$). Comparing this equation to \eqref{Zlem}, it is clear that we wish to establish that the last two terms are zero if $l<k+2$. Consider them in turn. The second term contains $d^k_{p,l-p-1}$, which by \eqref{dars} is zero if $p+l-p-1\leq k$, which it is if $l<k+2$. The third term contains $d^k_{q,p,l-p-q-2}$, which by \eqref{darst} equals zero as long as $q+p+l-p-q-2\leq k$, i.e. $l\leq k+2$, again implied by $l<k+2$.$\Box$\newpar
Therefore we may write:
\begin{equation}\label{Zlem2}
\bsym{\sigma}^k\bsym{\mc{Z}}=k(k+1)\xi^2_kx^k+\slim_{l=k+2}^y\bsym{\sigma}^k\bsym{f}_lx^l.
\end{equation}
Similarly to before, we substitute in $\psi^2_k$ \eqref{dxi2}, divide through by $x^{k+1}\mu$, rearrange \eqref{xi2e} and use \eqref{Zlem2} to get
\begin{equation}
\begin{split}
x\frac{d\psi^2_k}{dx}=&-2(k+1)\psi^2_k-k(k+1)\xi^2_kx^{-1}+k(k+1)\xi^2_k\frac{x^{-1}}{\mu}\\
&+\mu^{-1}\slim_{l=1}^{y-k-1}\bsym{\sigma}^k\bsym{f}_{l+k+1}x^l+\frac{1}{S}\frac{dS}{dx}\left(x\psi^2_k+k\xi^2_k\right).
\end{split}
\end{equation}
Again letting $\mu^{-1}=1+x^2\hat{\mu}$, using $S^{-1}=S_0^{-1}+x^2\hat{S}_C$, and substituting the expansion for the second Einstein equation \eqref{xds}, we obtain
\begin{equation}\label{dpsi22}
x\frac{d\psi^2_k}{dx}=-2(k+1)\psi^2_k+x\hat{\mc{J}}_{k}.\\
\end{equation}
This is now in the form we require. As a last step we must rewrite \eqref{dxi1} using the new variables \eqref{barpsi1}; this transformation does not significantly affect the structure of the equations, which become
\begin{equation}
\begin{split}
x\frac{d\xi^1_k}{dx}=&x\left(\bar{\psi}^1_k-\frac{1}{2(k+1)}\bsym{\sigma}^k\bsym{g}_{k+2}\right)\equiv x\hat{\mc{G}}_{k}.
\end{split}
\end{equation}
Hence, all our field equations are in the required form we need for them to apply to Theorem \ref{exdetheor}; and furthermore, it can be seen that the result \eqref{dpsi22} is indeed almost identical to the other transformed gauge field equation \eqref{xdpsik}, just as we expected from Lemma \ref{betaproptheta}.\newpar
So, to summarise, we have brought the field equations into the following forms:
\begin{equation}
\begin{array}{lll}
x\disfrac{d\chi}{dx}=-3\chi+x\hat{m}, & \,\, x\disfrac{d\xi^1_k}{dx}=x\hat{\mc{G}}_{k}, & \,\, x\disfrac{d\bar{\psi}^1_k}{dx}=-2(k+1)\bar{\psi}^1_k+x\hat{\mc{H}}_{2,k},\\
&&\\
x\disfrac{dS}{dx}=x^2\hat{S}_A, & \,\, x\disfrac{d\xi^2_k}{dx}=x\psi^2_k, & \,\, x\disfrac{d\psi^2_k}{dx}=-2(k+1)\psi^2_k+x\hat{\mc{J}}_{k}.\\
\end{array}
\end{equation}
It may be verified by substituting back that the functions $\hat{m}$, $\hat{S}_A$, $\hat{\mc{G}}_{k}$, $\hat{\mc{H}}_{2,k}$ and $\hat{\mc{J}}_{k}$ are all regular in all transformed field variables at $x=0$. At last, using Theorem \ref{exdetheor} gives us the following expansions at the origin:
\begin{equation}
\begin{array}{lll}
\chi(x)=O(x), & \,\, \xi^1_k(x)=\xi^1_{k,0}+O(x), & \,\, \bar{\psi}^1_k(x)=O(x),\\
&&\\
S(x)=S_0+O(x^2), & \,\, \xi^2_k(x)=\xi^2_{k,0}+O(x), & \,\, \psi^2_k(x)=O(x).\\
\end{array}
\end{equation}
By transforming our variables back, it can be checked that this gives the required behaviour for each of the field variables near the origin, completing the proof of Proposition \ref{lex0}.$\Box$

\subsection{Local existence of solutions at the event horizon}

Now we turn our attention to the field equations in the black hole case, i.e. at $r=r_h$ (for $r_h\neq0$). We assume the existence of a non-degenerate event horizon, so that $\mu(r_h)=0$ but $\mu'(r_h)>0$ is finite. Fortunately the black hole case is a great deal simpler than the previous soliton case. We begin with our proposition:

\begin{prop}\label{lexrh}
There exists an $2N-2$-parameter family of local solutions of the field
equations (\ref{EE1}, \ref{EE2}, \ref{YM1}, \ref{YM2}) near $r=r_h$ analytic in $r_h$, $\Lambda$, $\omega_{j,h}$, $\alpha'_{j,h}$ and $\rho=r-r_h$ such that
\begin{equation}
\begin{split}
\mu(r_h+\rho)&=\mu'_h\rho+O(\rho^2),\\
S(r_h+\rho)&=S_h+O(\rho),\\
\omega_j(r_h+\rho)&=\omega_{j,h}+O(\rho),\\
\alpha_j(r_h+\rho)&=\alpha'_{j,h}\rho+O(\rho^2),\\
\end{split}
\end{equation}
where $\mu'_h$ and $S_h$ are fixed by physicality requirements.
\end{prop}
\vsp
\textbf{Proof} Let us define a new independent variable \cite{BFM1994_static_spher_symm, MW1998_ex_sun, OK2002_gen_spher_EYM} $x=r-r_h$, and also define
some new dependent variables:
\begin{equation}
\begin{array}{ll}
\rho(x)=r, & \\
&\\
\lambda_1(x)=\disfrac{\mu(r)}{x},& \lambda_2(x)=S(r),\\
&\\
\psi_{1j}(x)=\disfrac{\alpha_{j}(r)}{x},&\psi_{2j}(x)=\omega_j(r),\\
&\\
\xi_{1j}(r)=\disfrac{\rho(x)^2}{S(r)}\disfrac{d\alpha_j}{dr},&\xi_{2j}(x)=\disfrac{\mu(r)}{x}\disfrac{d\omega_j}{dr}.\\
\end{array}
\end{equation}
The field equations take the form:
\begin{equation}\label{rhlocalex}
\begin{array}{ll}
x\disfrac{d\rho}{dx}=x, & x\disfrac{d\Lambda}{dx}=0,\\
&\\
x\disfrac{d\lambda_1}{dx}=-\lambda_1+\mathcal{F}_{h}+x\mathcal{G}_{1,h}, & x\disfrac{d\lambda_2}{dx}=x\mathcal{G}_{2,h},\\
&\\
x\disfrac{d\psi_{1j}}{dx}=-\psi_{1j}+\disfrac{\xi_{1j}\lambda_2}{\rho^2}, & x\disfrac{d\psi_{2j}}{dx}=\disfrac{x}{\lambda_1}\xi_{2j},\\
&\\
x\disfrac{d\xi_{1j}}{dx}=\disfrac{x}{\lambda_1\lambda_2}\psi_{2j}^2\left(\psi_{1j}-\psi_{1,j+1}\right), & x\disfrac{d\xi_{2j}}{dx}= -\xi_{2j}+\mathcal{P}_{j,h}+x\mathcal{H}_{j,h};\\
\end{array}
\end{equation}
where
\begin{equation}
\begin{split}
\mathcal{F}_{h}=&\frac{k}{\rho}+\frac{3\rho}{\Lambda^2}-\frac{1}{2\rho^3}\sum^N_{j=1}\left(\psi_{2j}^2-\psi_{2,j-1}^2-k\left(N+1-2j\right)\right)^2-\frac{1}{2\rho^3}\sum^N_{j=1}\xi_{1j}^2,\\
\mathcal{G}_{1,h}=&-\frac{\lambda_1}{\rho}-\frac{2}{\rho\lambda_1}\sum^N_{j=1}\xi_{2j}^2,\\
\mathcal{G}_{2,h}=&\frac{2\lambda_2}{\rho\lambda_1^2}\sum^{N-1}_{j=1}\xi_{2j}^2+\frac{1}{\lambda_1^2\lambda_2\rho}\sum^{N-1}_{j=1}\psi_{2j}\left(\psi_{1j}-\psi_{1,j+1}\right)^2,\\
\mathcal{H}_{j,h}=&-\frac{1}{4\lambda_1\lambda_2^2}\left(\psi_{1j}-\psi_{1,j+1}\right)^2-\frac{\xi_{2j}}{\lambda_2}\mathcal{G}_{2,h},\\
\mathcal{P}_{j,h}=&-\frac{1}{\rho^2}\psi_{2j}\left(k-\psi_{2j}^2+\frac{1}{2}\left(\psi_{2,j+1}^2+\psi_{2,j-1}^2\right)\right).\\
\end{split}
\end{equation}
To finally put the equations \eqref{rhlocalex} the required form, we let
\begin{equation}
\begin{array}{ccc}
\tilde{\psi}_{1j}=\psi_{1j}-\disfrac{\lambda_2\xi_{1j}}{\rho^2}, & \quad\tilde{\lambda}_1=\lambda_1-\mathcal{F}_{h}, & \quad\tilde{\xi}_{2j}=\xi_{2j}-\mathcal{P}_{j,h}.
\end{array}
\end{equation}
This makes the non-conforming equations take the form:
\begin{equation}\label{nonconrh}
x\frac{d\tilde{\lambda}_1}{dx}=-\tilde{\lambda}_1+x\tilde{\mathcal{G}}_{1,h}, \quad x\frac{d\tilde{\psi}_{1j}}{dx}=-\tilde{\psi}_{1j}+x\tilde{\mathcal{J}}_{j,h}, \quad x\frac{d\tilde{\xi}_{2j}}{dx}=-\tilde{\xi}_{2j}+x\tilde{\mathcal{H}}_{j,h},
\end{equation}
where
\begin{equation}
\begin{split}
\tilde{\mathcal{G}}_{1,h}=&\mathcal{G}_{1,h}-\disfrac{\partial\mathcal{F}_{h}}{\partial\rho}-\disfrac{\xi_{2j}}{\lambda_1}\disfrac{\partial\mathcal{F}_{h}}{\partial\psi_{2j}}-\disfrac{\psi_{2j}^2(\psi_{1j}-\psi_{1,j+1})}{\lambda_1\lambda_2}\disfrac{\partial\mathcal{F}_{h}}{\partial\xi_{1j}},\\
\tilde{\mathcal{H}}_{j,h}=&\mathcal{H}_{j,h}-\disfrac{\partial\mathcal{P}_{j,h}}{\partial\rho}-\disfrac{\partial\mathcal{P}_{j,h}}{\partial\psi_{2j}},\\
\tilde{\mathcal{J}}_{j,h}=&-\disfrac{\xi_{1j}}{\rho^2}\mathcal{G}_{2,h}-\disfrac{\psi_{2j}^2}{\lambda_1\rho^2}(\psi_{1j}-\psi_{1,j+1})+\disfrac{2\lambda_2\xi_{1j}}{\rho^3}.\\
\end{split}
\end{equation}
Hence, examining \eqref{rhlocalex} and \eqref{nonconrh}, we can use Theorem \ref{exdetheor} to show that there exist solutions to the equations (\ref{EE1}, \ref{EE2}, \ref{YM1}, \ref{YM2}) of the form
\begin{equation}
\begin{array}{llll}
\rho=r_h+O(x), && \\
&&\\
\tilde{\lambda}_1=O(x), & \,\, \psi_{1j}=\psi_{1j,h}+O(x), & \,\, \tilde{\xi}_{1j}=O(x),\\
&&\\
\lambda_2=\lambda_{2,0}+O(x), & \,\, \tilde{\psi}_{2j}=\tilde{\psi}_{2j,h}+O(x), & \,\, \xi_{2j}=O(x),\\
\end{array}
\end{equation}
with $\rho$, $\tilde{\lambda}_1$, $\lambda_2$, $\psi_{1j}$, $\tilde{\psi}_{2j}$, $\tilde{\xi}_{1j}$ and $\xi_{2j}$ all analytic in $x$, $r_h$, $\omega_j(r_h)$, $\alpha'_{j}(r_h)$, $\Lambda$, and $S(r_h)$. Transforming back to our original variables gives us the correct behaviour and analyticity. When we fix $r_h$ and $\Lambda$ and choose $S_h$ such that $S\rar1$ as $r\rar\infty$, this gives the expected $2N-2$ parameters. $\Box$\newpar
We have thus proven existence of solutions to the field equations for a black hole in some neighbourhood of the event horizon $r=r_h$, satisfying the boundary conditions \eqref{exprh}. 

\subsection{Local existence of solutions at infinity}
\label{exinf}

Now we prove existence locally as $r\rar\infty$, which applies to both black hole and soliton solutions. We note that, as in the adS spherically symmetric case, it is relatively easy to prove existence here and we need only go to first order in the field variables, unlike in the asymptotically flat case where higher order terms were needed and the analysis was much more involved \cite{OK2002_gen_spher_EYM, HPK1994_analysis_sun}.
\begin{prop}
\label{lexinf}
There exists an $2N$-parameter family of local solutions of the
field equations (\ref{EE1}, \ref{EE2}, \ref{YM1}, \ref{YM2}) near $r=\infty$, analytic in $\Lambda$, $\omega_{j,\infty}$, $M$ and $r^{-1}$ such that
\begin{equation}
\begin{split}
\mu(r)&=k-\frac{2M}{r}-\frac{\Lambda r^2}{3}+O\left(r^{-2}\right),\\
S(r)&=S_\infty+O\left(r^{-4}\right),\\
\omega_j(r)&=\omega_{j,\infty}+\frac{c_j}{r}+O\left(r^{-2}\right).\\
\alpha_j(r)&=\alpha_{j,\infty}+\frac{d_j}{r}+O\left(r^{-2}\right).\\
\end{split}
\end{equation}
\end{prop}
\vsp
\textbf{Proof} We transform our independent variable to $x=r^{-1}$, and introduce new variables \cite{BFM1994_static_spher_symm, HPK1994_analysis_sun, BW2008_sun_ex}:
\begin{equation}
\begin{array}{lll}
\lambda_1(x)=2m(r), & \,\,\psi_{1j}(x)=\alpha_j(r), & \,\,\xi_{1j}(x)=r^2\disfrac{d\alpha_j}{dr},\\
&&\\
\lambda_2(x)=S(r), & \,\,\psi_{2j}(x)=\omega_j(r), & \,\, \xi_{2j}(x)=r^2\disfrac{d\omega_j}{dr}.\\
\end{array}
\end{equation}
Then the field equations take the form:
\begin{equation}
\begin{array}{lll}
x\disfrac{d\lambda_1}{dx}=x\mathcal{G}_{\infty,1}, & \quad x\disfrac{d\psi_{1j}}{dx}=-x\xi_{1j}, & 
\quad x\disfrac{d\xi_{1j}}{dx}=x\mathcal{H}_{\infty,1j},\\
&&\\
x\disfrac{d\lambda_2}{dx}=x^4\mathcal{G}_{\infty,2}, & \quad x\disfrac{d\psi_{2j}}{dx}=-x\xi_{2j}, & \quad x\disfrac{d\xi_{2j}}{dx}=x\mathcal{H}_{\infty,2j},\\
&&\\
x\disfrac{d\Lambda}{dx}=0; &&\\
\end{array}
\end{equation}
where
\begin{equation}
\begin{split}
\mathcal{F}_{\infty}=&-\disfrac{\Lambda}{3}+kx^2-x^3\lambda_1=\mu x^2,\\
\mathcal{G}_{\infty,1}=&-\disfrac{1}{\lambda_2^2}\sum^N_{j=1}\xi_{1j}^2-\disfrac{1}{2\mathcal{F}_\infty\lambda_2^2}\sum^{N-1}_{j=1}\psi_{2j}^2\left(\psi_{1j}-\psi_{1,j+1}\right)^2,\\
&-2\mathcal{F}_\infty\sum^{N-1}_{j=1}\xi_{2j}^2-\frac{1}{2}\sum^N_{j=1}\left(\psi_{2j}^2-\psi_{2,j+1}^2-k(N+1-2j)\right)^2,\\
\mathcal{G}_{\infty,2}=&-\disfrac{1}{2\mathcal{F}_\infty^2\lambda_2}\sum^{N-1}_{j=1}\psi_{2j}^2(\psi_{1j}-\psi_{1,j+1})-2\lambda_2\sum^{N-1}_{j=1}\xi_{2j}^2,\\
\mathcal{H}_{\infty,1j}=&\disfrac{1}{\mathcal{F}_\infty}\left(\psi_{2j}^2(\psi_{1j}-\psi_{1,j+1})-\psi_{2,j-1}^2(\psi_{1,j-1}-\psi_{1j})\right)+\disfrac{x^3\xi_{1j}\mathcal{G}_{\infty,2}}{\lambda_2},\\
\mathcal{H}_{\infty,2j}=&\disfrac{\psi_{2j}(\psi_{1j}-\psi_{1,j+1})}{4\mathcal{F}_\infty^2\lambda_2^2}+\disfrac{\psi_{2j}}{\mathcal{F}_\infty}\left(k-\psi_{2j}^2+\frac{1}{2}\left(\psi_{2,j-1}^2+\psi_{2,j+1}^2\right)\right),\\
&+\disfrac{x\xi_{2j}(3x\lambda_1-2k)}{\mathcal{F}_\infty}+\disfrac{x^3\xi_{2j}\mathcal{G}_{\infty,1}}{\mathcal{F}_\infty}-\disfrac{x^3\xi_{2j}\mathcal{G}_{\infty,2}}{\lambda_2}.\\
\end{split}
\end{equation}
Since $1/\mu$ is at least of order $x^2$ as $x\rightarrow 0$, it can be observed that all of these polynomials are non-singular as $x\rightarrow 0$. Therefore using Theorem \ref{exdetheor}, we have solutions to these equations with the following asymptotic behaviour:
\begin{equation}
\begin{array}{lll}
\lambda_1(x)=\lambda_{1,0}+O(x), \quad & \psi_{1j}=\psi_{1j,0}+O(x), \quad & \xi_{1j}=\xi_{1j,0}+O(x),\\
\lambda_2(x)=\lambda_{2,0}+O(x^4), \quad & \psi_{2j}=\psi_{2j,0}+O(x), \quad & \xi_{2j}=\xi_{2j,0}+O(x).\\
\end{array}
\end{equation}
Therefore we have proven local existence of solutions at infinity, and Theorem \ref{exdetheor} confirms that the functions exhibit the required behaviour near infinity \eqref{expinf}. Also, by rescaling the time co-ordinate in the metric we can fix $S_\infty=1$ so that the space-time is asymptotically topological adS -- this satisfies the boundary conditions \eqref{expinf}, and the field variables are thus analytic in $M$, $x$, $\omega_{j,\infty}$, $\alpha_{j,\infty}$, $c_j$, $d_j$, and $\Lambda$. Finally, fixing $r_h$ and $\Lambda$, we get the $4N-3$ parameter family of solutions that we were expecting.$\Box$

\section{Global aspects of non-trivial solutions}
\label{gloexsec}

Having proven local existence at the various boundaries $r=0$, $r=r_h$ and $r\rar\infty$, we now turn our attention to proving that those solutions may be patched together into global solutions, i.e. solutions which begin at $r=r_h$ for black holes ($r=0$ for solitons) and remain regular throughout the range $r_h\leq r\leq \infty$ for black holes ($0\leq r\leq \infty$ for solitons). First though we prove a more minor proposition concerning $\mc{E}_j$, which it is worth including nonetheless.

\begin{prop}\label{emon}
$\mc{E}_j(r)$ is monotonic for all $j\in\{1,...,N-1\}$ -- i.e., $\mc{E}_j(r)\mc{E}'_j(r)>0\,\,\forall j\in\{1,...,N-1\}$ and $\forall r$ such that $r_h<r<\infty$ for black hole solutions, $0<r<\infty$ for soliton solutions.
\end{prop}
\textbf{Proof} Inspired by Section \ref{lexboundsec}, we may write the Yang-Mills equation \eqref{YM1E} in terms of the matrix $A^i_j$ \eqref{matA} as
\begin{equation}
\mu S\slim_{i=1}^{N-1}\left(\frac{r^2}{S}\mc{E}'_i\right)'\kron^i_j=\slim_{i=1}^{N-1}A^i_ju_i^2\mc{E}_i.
\end{equation}
Post-multiplying by $v^j_C$ (for some integer $1\leq C\leq N-1$), using (\ref{betathetaUE}, \ref{Abethe}) and summing over $j$ gives
\begin{equation}\label{mon2}
\mu S\slim_{i=1}^{N-1}\left(\frac{r^2}{S}\mc{E}'_i\right)'v^i_C=\slim_{i=1}^{N-1}C(C+1)u_i^2\mc{E}_iv^i_C.
\end{equation}
It is not important what the value of $C(C+1)$ is -- we merely note that it is positive (as an eigenvalue of $A^i_j$) and hence we define $\tilde{C}\equiv C(C+1)>0$. Now what we wish to prove is that the above relationship holds for each summand in $i$, and hence that we can diagonalise the system, decoupling the equations. We rewrite \eqref{mon2} to clarify the situation:
\begin{equation}\label{mon3}
\slim_{i=1}^{N-1}\left[\mu S\left(\frac{r^2}{S}\mc{E}'_i\right)'-\tilde{C}u_i^2\mc{E}_i\right]v^i_C=\bsym{0}
\end{equation}
where $\bsym{0}=(0,0,...,0)^T$, the zero vector of length $N-1$. We now consider the system of vectors $v^i_C$ as being labelled by $i$ and indexed by $C$: i.e., we take the set of vectors $\{v^i_C\}\equiv\{v^1_C,v^2_C,...,v^{N-1}_C\}$. Elementary linear algebra tells us that if we can prove that $\{v^i_C\}$ forms a linearly independent set of $N-1$ vectors of length $N-1$, it is a basis, and hence \eqref{mon3} shows that the equation must hold for each value of $i$. Considering that the $v^i_C$ are defined using Hahn polynomials it should be no surprise that this holds in our case, due to inherited orthogonality relations and so on; but we shall show the direct proof in any case. In other words, we would prove that given $N-1$ coefficients $\nu_i$,
\begin{equation}\label{mon4}
\slim_{i=1}^{N-1}\nu_iv^i_C=\bsym{0}\iff \nu_i=0\,\,\forall i,
\end{equation}
where for now we assume nothing about the coefficients $\nu_i$. It is simple to show that $\nu_i=0\,\,\forall i$ is a sufficient condition for $\slim_{i=1}^{N-1}\nu_iv^i_C=\bsym{0}$. To show that it is necessary, we need to derive an analogous orthogonality relationship to \eqref{vwnorm}. Comparing the results (\ref{evec}, \ref{vwnorm}, \ref{dk}) from K\"{u}nzle \cite{HPK1994_analysis_sun} (see lemma 2) with the orthogonality and dual orthogonality relationships for the Hahn polynomials \cite{SK1961_Hahn}, which are 
\begin{equation}\label{orth}
\begin{split}
\slim_{x=0}^{N-1}Q_n(x)Q_m(x)\rho(x)&=\frac{1}{\pi_n}\kron^n_m,\\
\slim_{n=0}^{N-1}Q_n(x)Q_n(y)\pi_n&=\frac{1}{\rho(x)}\kron^x_y,
\end{split}
\end{equation}	
we find that in our case,
\begin{equation}
\pi_n=d_n^{-1}N(N-1),\quad\quad\rho(x)=\frac{x+1}{N(N-x-1)}.
\end{equation}
Using these with \eqref{orth} and the above results from K\"{u}nzle it is possible to show that
\begin{equation}\label{dualorth}
\slim_{C=1}^{N-1}d_C^{-1}\sigma^C_kv^i_C=\bsym{\delta}^i_k.
\end{equation}
The analogy to \eqref{vwnorm} is obvious. So, to prove \eqref{mon4}, we multiply the left-hand side through by $\sigma^C_kd_C^{-1}$ (for some $1\leq k\leq N-1$), sum over $C$, and use \eqref{dualorth} to show that
\begin{equation}\label{mon6}
\begin{split}
&\slim_{C=1}^{N-1}\slim_{i=1}^{N-1}\nu_id_C^{-1}\sigma^C_kv^i_C=0\\
\implies&\slim_{i=1}^{N-1}\nu_i\bsym{\delta}^i_k=0\\
\implies&\nu_k=0.
\end{split}
\end{equation}
Since $k$ was arbitrary, we have shown that $\nu_j=0\,\,\forall j$. Therefore, since $\nu_i$ is arbitrary, we have proven that $\{v^1_C,v^2_C,...,v^{N-1}_C\}$ is a basis. Finally, to agree with \eqref{mon3} we choose
\begin{equation}\label{mon5}
\nu_j=\mu S\left(\frac{r^2}{S}\mc{E}'_j\right)'-\tilde{C}u_j^2\mc{E}_j=0
\end{equation}
(no sum on $j$), and it is clear that we have diagonalised this system to
\begin{equation}
\left(\frac{r^2}{S}\mc{E}'_j\right)'=\frac{\tilde{C}}{\mu S}u_j^2\mc{E}_j\,\,\,\forall j,
\end{equation}
at least up to a factor of an unknown positive coefficient $\tilde{C}$, which may in general depend on $j$. Thus, at least for the purposes of this proof, the system is decoupled.

For ease of notation we define $r_0=r_h$ for black holes and $r_0=0$ for solitons. We recall that $\mc{E}_j(r_0)=0$ and $\mc{E}'_j(r_0)\neq 0$ in general. Therefore, noting that $\tilde{C}$, $r^2$, $\mu$, $S$ and $u_i^2$ are all positive, then by integrating both sides it is easy to establish that for $r>r_0$, $\mc{E}'_j(r)$ has the same sign as $\mc{E}_j(r)$ -- i.e. $\mc{E}_j(r)$ is monotonic; increasing (decreasing) if $\mc{E}'_j(r_0)>0$ ($\mc{E}'_j(r_0)<0$).$\Box$

\vsp
\textbf{Comment} We briefly note that we unfortunately cannot transfer this result into telling us something similar about $\alpha_j(r)$, since the $\mc{E}_j$ are the \textit{differences} between each pair of successive functions $\alpha_j(r)$. In the case of $\essu$ alone, there is only one $\mc{E}_j\equiv\mc{E}$ and only one independent EGF $\alpha_1\equiv\alpha$, and since $\alpha_1+\alpha_2=0$ for $\essu$, we find $\mc{E}=\alpha_1-\alpha_2=2\alpha_1=-2\alpha_2=2\alpha$. What is more, $\alpha$ is monotonic in the $\essu$ case \cite{NW2012_dyon}. Therefore, both $\alpha_j(r)$ are monotonic there, and hence for embedded solutions, $\mc{E}_j\equiv2\alpha$ are monotonic $\forall j$. So this result does imply at least that any $\essu$ embedded solutions will also have all $\alpha_j$ monotonic; though we note that due to tracelessness and the transform itself, various $\alpha_j$ will be positive (negative) and monotonically increasing (decreasing) for $r>r_h$ for black holes or $r>0$ for solitons, with at least one $\alpha_j$ of each sign.

\subsection{Global regularity for $\mu(r)>0$}
\label{gloregsec}

We now need to prove that the local solutions that we found will remain regular if we begin with their initial conditions and integrate out arbitrarily far. We again let $r_0=r_h$ for black holes, and $r_0=0$ for solitons, so that we are discussing regularity in the range $\hat{R}\equiv(r_0,\infty)$. Also, note that we require the metric function $\mu(r)>0$ for all $r\in\hat{R}$; but we expect $\hat{R}$ to correspond to the external region of a black hole (and $\mu(r)>0\,\,\forall r\in\hat{R}$ for solitons), therefore taking $\mu(r)>0\,\,\forall r\in\hat{R}$ is a natural physical requirement. We state the proposition:

\begin{prop}\label{gloreg}
As long as $\mu(r)>0$, the field equations (\ref{EE1}, \ref{EE2}, \ref{YM1}, \ref{YM2}) remain regular in all field variables throughout the range $r_h<r<\infty$ for black holes, and $0<r<\infty$ for solitons.
\end{prop}
\textbf{Proof} Take some $r_1>r_0$, and define intervals $\mc{Q}=(r_0,r_1)$ and $\bar{\mc{Q}}=(r_0,r_1]$. Since we have proven existence in some neighbourhood of the boundary points represented by $r=r_0$, our strategy is to assume that the field variables remain regular on $\mc{Q}$, and use the field equations to prove that they continue to remain regular on $\bar{\mc{Q}}$ (i.e. at $r=r_1$), as long as $\mu(r)>0\,\,\forall r\in\bar{\mc{Q}}$, and thus we may integrate the field equations out arbitrarily far and into the asymptotic regime.

We first note that all the terms on the right-hand sides of the Einstein equations \eqref{EE1} and \eqref{EE2} are non-negative, and therefore all of those terms are bounded below by 0 and above by the left-hand sides. This also implies that for non-trivial solutions (with non-constant $m(r)$ and $S(r)$), $m'(r)>0$ i.e. that $m(r)$ is monotonically increasing (since $m(r_0)>0$); and that $S'S^{-1}>0$ i.e. $S'(r)S(r)>0$, and since we can fix $S_0$ to be positive, then $S(r)$ is also monotonically increasing. Finally, an important upper bound is given by $\mu(r_1)>0$, which implies
\begin{equation}
2m(r_1)<kr_1+\frac{r_1^3}{\ell^2}.
\end{equation}
The right-hand side of this is positive (recalling the minimum event horizon radius for $k=-1$ \eqref{minrhk-1}), so that $m(r)$ is regular on $\bar{\mc{Q}}$ and therefore so is $\mu(r)$. Now we consider \eqref{EE1}: due to the preceding comments, we may write
\begin{equation}\label{mrgr}
2m'(r)\geq 2\mu G+\frac{\zeta}{2\mu S^2}.
\end{equation}
Now $\mu(r)$ must have a minimum value on $\bar{\mc{Q}}$, so define
\begin{equation}
\mu_{min}=\min\{\mu(r):r\in\bar{\mc{Q}}\}. 
\end{equation}
Then, integrating both sides of \eqref{mrgr}, we obtain
\begin{equation}\label{Sbound}
\frac{2[m(r_1)-m(r_0)]}{\mu_{min}}\geq\int\limits_{r_0}^{r_1}\left(2G+\frac{\zeta}{2\mu^2 S^2}\right)dr,
\end{equation}
and direct integration of \eqref{EE2} gives us that $\ln|S(r)|$ and therefore $S(r)$ is bounded in $\bar{\mc{Q}}$. 

Therefore, each of the integrals on the right-hand side of \eqref{Sbound} must also be bounded. So, using the Cauchy-Schwartz inequality:
\begin{equation}
\begin{split}
\slim_{j=1}^{N-1}\left[\omega_j(r_1)-\omega_j(r_0)\right]^2=\slim_{j=1}^{N-1}\left(\int\limits_{r_0}^{r_1}\omega_j'(r)dr\right)^2&\leq(r_1-r_0)\int\limits_{r_0}^{r_1}\slim_{j=1}^{N-1}\omega_j'^2(r)dr\\
&\leq (r_1-r_0)\int\limits_{r_0}^{r_1}Gdr,\\
\end{split}
\end{equation}
and since the left-hand side is a sum of positive terms and the right-hand side is bounded above by \eqref{Sbound}, we can say that $\omega_j(r)$ is bounded on $\bar{\mc{Q}}$ for all $j\in\{1,...,N-1\}$.

In a similar manner we directly integrate \eqref{EE1} and consider only the first term on the right-hand side, which in turn is bounded above by the bounds on $m(r)$:
\begin{equation}
[m(r_1)-m(r_0)]\geq\int\limits_{r_0}^{r_1}\frac{r^2}{4S^2}\slim_{j=1}^{N-1}\alpha_j'^2(r)dr.
\end{equation}
Extracting factors of bounded (positive) functions, it is elementary to show that
\begin{equation}
\frac{4S(r_1)^2}{r_0^2}[m(r_1)-m(r_0)]\geq\int\limits_{r_0}^{r_1}\slim_{j=1}^{N-1}\alpha_j'^2(r)dr.
\end{equation}
However, again using Cauchy-Schwartz, we find
\begin{equation}
\slim_{j=1}^{N}\left[\alpha_j(r_1)-\alpha_j(r_0)\right]^2=\slim_{j=1}^{N}\left(\int\limits_{r_0}^{r_1}\alpha_j'(r)dr\right)^2\leq(r_1-r_0)\int\limits_{r_0}^{r_1}\slim_{j=1}^{N}\alpha_j'^2(r)dr,
\end{equation}
and so likewise $\alpha_j(r)$ is regular on $\bar{\mc{Q}}$ for all $j\in\{1,...,N\}$, which in turn means that so is $\mc{E}_j(r)$, by definition.\newpar
Finally we examine the Yang-Mills equations in the forms \eqref{YM2} and \eqref{YM1E}. We begin by rewriting them as:
\begin{align}
\left(\frac{r^2}{S}\mc{E}_j'\right)'&=\frac{1}{\mu S}\mc{Z}_j,\label{YM1glo}\\
\left(\mu S\omega_j'\right)'&=\frac{-SW_j\omega_j}{r^2}-\frac{1}{4\mu S}\omega_j\mc{E}_j^2.\label{YM2glo}
\end{align}
Begin with (\ref{YM2glo}). Note that all terms on the right-hand side are bounded on $\bar{\mc{Q}}$, so we can immediately write
\begin{equation}
\mu(r_1)S(r_1)\omega'_j(r_1)=\mu(r_0)S(r_0)\omega'_j(r_0)-\int\limits_{r_0}^{r_1}\left(\frac{SW_j\omega_j}{r^2}+\frac{1}{4\mu S}\omega_j\mc{E}_j^2\right)dr,
\end{equation}
and therefore the left-hand side is bounded, and so $\omega'_j(r)$ is regular on $\bar{\mc{Q}}$ for all $j\in\{1,...,N-1\}$.\newpar
Finally, taking (\ref{YM1glo}), the same is true: we may write
\begin{equation}
\frac{r_1^2}{S(r_1)}\mc{E}'_j(r_1)=\frac{r_0^2}{S(r_0)}\mc{E}'_j(r_0)+\int\limits_{r_0}^{r_1}\frac{1}{\mu S}\mc{Z}_jdr,
\end{equation}
showing that since the right-hand side is bounded, the left-hand side will be too. Therefore $\mc{E}'_j(r)$ is also bounded on $\bar{\mc{Q}}$ for all $j\in\{1,...,N-1\}$. Thus Proposition \ref{gloreg} is proven.$\Box$

\subsection{Asymptotic behaviour}
\label{asymptotics}

However, our preparations are still not quite complete. A big difference between the $\Lambda=0$ and $\Lambda<0$ EYM cases is the dearth of solutions in the former case and the abundance in the latter, and this is entirely due to the different geometry of the manifold in the limit $r\rar\infty$. Therefore, it is important to examine the behaviour of the equations in the asymptotic limit.

We begin with the Yang-Mills equations in the same form as in Section \ref{gloregsec}:
\begin{align}
\left(\frac{r^2}{S}\mc{E}_j'\right)'&=\frac{1}{\mu S}\mc{Z}_j,\label{YM1asym}\\
\left(\mu S\omega_j'\right)'&=\frac{-SW_j\omega_j}{r^2}-\frac{1}{4\mu S}\omega_j\mc{E}_j^2.\label{YM2asym}
\end{align}
First we let $r\rar\infty$, so that $\mu(r)\rar\frac{r^2}{\ell^2}$ and $S(r)\rar 1$, and then we make the co-ordinate change $r=\ell^{-1}\tau^{-1}$ to convert the asymptotic field equations into autonomous form. This yields
\begin{equation}\label{asym}
\begin{split}
\frac{d^2\mc{E}_j}{d\tau^2}&=\ell^4\mc{Z}_j,\\
\frac{d^2\omega_j}{d\tau^2}&=-\ell^4\left(W_j+\frac{\ell^2}{4}\mc{E}_j^2\right)\omega_j.
\end{split}
\end{equation}
It may be seen that although these equations are clearly autonomous, they are not scale-invariant as they contain a reference to $\ell$; however this is not a problem as we are not interested in the limit $\ell\rar\infty$ (i.e. $|\Lambda|\rar0$), and the equations remain well-behaved if we take $\ell$ arbitrarily small (which we shall consider later). Suffice to say that equations \eqref{asym} are easily solved to find the critical points $\bar{\omega}_{j}$ and $\bar{\mc{E}}_{j}$ -- the relevant equations are
\begin{align}
2\bar{\omega}_{j}^2\bar{\mc{E}}_{j}-\bar{\omega}_{j-1}^2\bar{\mc{E}}_{j-1}-\bar{\omega}_{j+1}^2\bar{\mc{E}}_{j+1}&=0,\label{crit1}\\
\left(\frac{\ell^2}{4}\bar{\mc{E}}_j^2+\left(k-\bar{\omega}^2_j+\frac{1}{2}\left(\bar{\omega}_{j-1}^2+\bar{\omega}_{j+1}^2\right)\right)\right)\bar{\omega_j}&=0;\label{crit2}
\end{align}
and certain elementary solutions are given as follows: 
\begin{enumerate}[(i)]
\item For all values of $k$, we get the solution $\bar{\omega}_j=0$, $\bar{\mc{E}}_j$ arbitrary;
\item For $k=1$ only, we get the solution $\bar{\omega}_j=\pm\sqrt{j(N-j)}$, $\bar{\mc{E}}_j=0$;
\end{enumerate}

where (i) is a metastable centre in the phase space and (ii) are saddle points, as we may expect \cite{BW2008_sun_ex}. Note that these solutions are independent of the value of $\ell$. Because (\ref{crit1}, \ref{crit2}) are non-linear equations, there may potentially be other critical points; but it can at least be observed that for each critical point, the given value for $\bar{\omega}_j$ solves the system (\ref{crit1}, \ref{crit2}) if and only if the given value for $\bar{\mc{E}}_j$ solves (\ref{crit1}, \ref{crit2}). 

We notice that the study of the solutions locally as $r\rar\infty$ implied no such constraints on the parameters. However, this is due to our choice of autonomous parameter, where $\tau\propto r^{-1}$. Similar to the purely magnetic adS cases \cite{BW2008_sun_ex, JEB2014_top_ex}, this means that if we consider our solution as a trajectory in the phase space of the $4N-4$-dimensional system 
\begin{equation}
\left(\omega_j(r),\disfrac{d\omega_j}{dr},\mc{E}_j(r),\disfrac{d\mc{E}_j}{dr}\right),
\end{equation}
and then consider the transform to the variable $\tau$, then we can see that integration over the range $r\in[r_1,\infty)$ becomes integration over $\tau\in[0,\tau_1]$, and if $r_1$ is large enough to be considered asymptotic then the corresponding trajectory in terms of $\tau$ will be very short. Hence the solution will not move all the way along the trajectory, and therefore the values of the gauge field at infinity are unconstrained. We can compare this to the flat space case, in which there is a very different situation: the parameter used there must be such that $\tau\propto\ln r$, hence $(r_1,\infty)\mapsto(\tau_1,\infty)$, and so in that case every solution must proceed to the end of its trajectory. This is the reason for the scarcity of solutions in the case $\Lambda=0$, where in the purely magnetic case it is proven that solutions can only be found for certain discrete values of the gauge field parameters at infinity.

\section{Global existence of non-trivial solutions}
\label{gloexsection}

Now we have all the machinery we need to construct global solutions which are genuinely non-trivial, i.e. solutions which do not appear in our list in Section \ref{triv}. In this penultimate section, we present arguments that allow us to piece together the local solutions we found at the boundaries in Section \ref{lexboundsec} using the global regularity results in Sections \ref{gloexsec}, and thus describe global non-trivial solutions to the field equations (\ref{EE1}, \ref{EE2}, \ref{YM1}, \ref{YM2}). We consider existence in several regimes, including in the limit of $|\Lambda|\rar\infty$.

\subsection{Existence of non-trivial solutions near existing solutions}

The crux of the proof of global existence is in the following powerful proposition, the essence of which is the proof that dyonic solutions to the field equations exist in open sets of the parameter space:

\begin{prop}\label{globex}
Assume we have an existing solution to the dyonic $\sun$ topological field equations (\ref{EE1}, \ref{EE2}, \ref{YM1}, \ref{YM2}), where each MGF $\omega_j(r)$ has $R_j$ nodes and gauge fields have initial values $\{\omega_{jh}, \mc{E}'_{jh}\}$ for topological black holes or $\{\breve{\beta}_{k}, \breve{\theta}_{k}\}$ for spherical solitons. Then all initial gauge field values in some sufficiently small neighbourhood of the existing solutions will also give an $\sun$ dyonic topological solution to the field equations (a spherical solution for solitons) in which each MGF $\omega_j(r)$ has $R_j$ nodes.
\end{prop}
\vsp
\textbf{Proof} Assume we know of an existing solution to the dyonic $\sun$ topological field equations, where each $\omega_j(r)$ has $R_j$ nodes. For black holes we have initial conditions $\{\omega_{jh}\neq0, \mc{E}'_{jh}\neq0\}$, and for solitons, $\{\breve{\beta}_{k}\neq0, \breve{\theta}_{k}\neq0\}$ (in the general case). From these initial conditions, Proposition \ref{gloreg} and Section \ref{asymptotics} it show that as long as $\mu(r)>0$ we may integrate this solution out arbitrarily far into the asymptotic regime to obtain a solution which will satisfy the boundary conditions as $r\rar\infty$. For the rest of the argument, we assume that $\ell$ and $r_h$ are fixed (where $r_h=0$ for solitons); that each MGF $\omega_j$ has $R_j$ nodes; and that again, $r_0=r_h$ for black holes and $r_0=0$ for solitons.

From the local existence results (Propositions \ref{lex0}, \ref{lexrh} and \ref{lexinf}), we know that for any set of initial values there are solutions locally near the event horizon (for a black hole, or the origin for a soliton), and that solutions are analytic in their choice of initial conditions. For an existing dyonic $\sun$ solution, it must be true that $\mu(r)>0$ for all $r\in[r_0,\infty)$. So, by analyticity, the nearby dyonic $\sun$ solution will also have $\mu(r)>0$ for all $r\in[r_0,r_c]$ for some $r=r_c$ with $r_0<r_c<\infty$. By Proposition \ref{gloreg}, this nearby solution will also be regular on $[r_0,r_c]$.

Now, choose some $r_1 >> r_0$ such that for the existing solution, $m(r_1)/r_1<<1$. Let $\{\check{\omega}_{jh}, \check{\mc{E}}'_{jh}\}$ (for a black hole solution, or $\{\check{\beta}_{k}, \check{\theta}_{k}\}$ for a soliton) be a different set of initial conditions at $r=r_0$ in some neighbourhood of the existing solution, and let $\check{m}(r)$ be the mass function of that solution. By analyticity (as above), these will also be regular on $[r_0,r_1]$ -- i.e. $\mu(r)>0$ on this interval -- and the MGFs $\omega_j$ will each have $R_j$ nodes.

Also it is then the case that $\check{m}(r_1)/r_1<<1$, and since $r_1 >> r_0$ we consider this the asymptotic regime. Provided $r_1$ is large enough (and hence $\tau_1$ is very small), the solution will not move very far along its phase plane trajectory as $r_1\rar\infty$. Therefore $\check{m}(r)/r$ remains small and the asymptotic regime remains valid. Therefore the solution will remain regular.$\Box$

\vsp
\textbf{Comment} We note that the set of generally dyonic solutions trivially includes the set of purely magnetic solutions, simply by choosing those ones for which $\mc{E}_j\equiv\alpha_j\equiv0\,\,\forall j$. Given then that this result also applies to existing topological solutions, $\essu$ dyonic solutions and purely magnetic $\sun$ solutions, this is a really rather general result that we can use to prove existence in a number of regimes, using these established solutions \cite{VdBR_su2_top, JEB2014_top_ex, NW2012_dyon, BW2008_sun_ex}. We also recall that we are most interested in solutions for which the MGFs have no nodes, so this is the kind of solution we look for here.

We sum up this subsection in the following theorem, which uses Proposition \ref{globex} and known trivial solutions and solutions from previous research to generate nearby non-trivial solutions. A similar structure of argument will apply in each case, so we choose to use a shorthand for brevity and clarity: $X\rrar Y$ will be shorthand for ``the existence of solution $X$ implies the existence of the nearby solution $Y$ using Proposition \ref{globex}". (For a longhand version of the argument structure, see Theorem 8 in our previous work \cite{JEB2014_top_ex}). Note that our Proposition \ref{globex} relies on solutions being analytic \textit{in some neighbourhood} of existing solutions, so while that guarantees that nodeless solutions will have nearby nodeless solutions, it also means that if we use a purely magnetic solution as our `trivial' solution, we will get a neighbouring solution for which all the $\mc{E}_j$ (and hence $\alpha_j$, due to their zero sum) will be \textit{small}; and if we use an $\essu$ embedded solution we get all $\alpha_j$ \textit{monotonic}. We have tried to highlight these properties in each case for the sake of completeness.

\begin{thr}\label{regimes}
\textbf{Regimes of existence of non-trivial nodeless solutions} 

Non-trivial global solutions exist to the field equations (\ref{EE1}, \ref{EE2}, \ref{YM1}, \ref{YM2}) in the following list of regimes found nearby existing solutions.
\end{thr}

\begin{itemize}
\item For strictly topological black holes (i.e. with $k\neq1$):

\begin{enumerate}
\item Topological $\essu$ purely magnetic solutions \cite{VdBR_su2_top}, all of which are nodeless\\
$\rrar$ Nearby non-trivial $\essu$ topological dyonic solutions, all of which are nodeless due to analyticity, and for which the only independent EGF $\alpha$ is monotonic and small -- and hence embedded $\sun$ topological dyonic solutions based on these (by \eqref{embed})\\
$\rrar$ Non-trivial (i.e. non-embedded) solutions nearby embedded $\sun$ topological dyonic solutions, again all of which are nodeless due to analyticity, and for which all $\alpha_j$ are monotonic and small.

\item Topological $\sun$ purely magnetic solutions \cite{JEB2014_top_ex}, some of which are nodeless\\
$\rrar$ Nearby non-trivial dyonic $\sun$ solutions, some of which are nodeless and for which all $\alpha_j$ are small, and not necessarily monotonic.
\end{enumerate}

\item For spherically symmetric black holes \ul{and} solitons (i.e. for $k=1$):

\begin{enumerate}\setcounter{enumi}{2}

\item Non-trivial dyonic $\essu$ solutions \cite{NW2012_dyon}, some of which are nodeless, imply the existence of $\sun$ embedded dyonic solutions \eqref{embed} \\
$\rrar$ Non-trivial (i.e. non-embedded) solutions nearby $\sun$ embedded dyonic solutions, some of which are nodeless, and for which $\alpha_j$ are all monotonic but otherwise quite general.

\item Non-trivial purely magnetic $\sun$ solutions \cite{BW2008_sun_ex}, some of which are nodeless\\
$\rrar$ Nearby non-trivial $\sun$ dyonic solutions, some of which are nodeless, and for which all $\alpha_j$ are small, and not necessarily monotonic.

\item For black holes only: The $\sun$ SadS purely magnetic solution (see Section \ref{triv}), which is nodeless\\
$\rrar$ Nearby non-trivial solutions all of which are nodeless, and for which all $\alpha_j$ are small and not necessarily monotonic.

\item For solitons only: The $\sun$ pure adS solution (see Section \ref{triv}), which is nodeless\\
$\rrar$ Nearby non-trivial $\sun$ dyonic solutions, all of which are nodeless, and for which all $\alpha_j$ are small and not necessarily monotonic.\\

\end{enumerate}

\end{itemize}

\subsection{Existence of solutions as $|\Lambda|\rar\infty$}
\label{l0exsec}

So far, we have proven the global existence of non-trivial solutions for any fixed value of the cosmological constant $\Lambda<0$. However, we see that numerical results in the purely magnetic case \cite{BWH2008_abund_stable, BW2008_sun_ex} imply that for fixed $\Lambda$, the gauge field parameter space in which we find nodeless MGFs shrinks as $N$ increases. However, we also find that if we make $|\Lambda|$ large enough, \textit{all} solutions we find have nodeless MGFs. In addition to this, we note that in the case of purely magnetic $\essu$ \cite{EW1999_su2_ex} and $\sun$ \cite{EB+EW_stab_sun} solutions, $|\Lambda|\rar\infty$ emerges analytically as a condition of stability in the gravitational sector.

We have not been able to prove that all solutions for $|\Lambda|\rar\infty$ are nodeless, but we could at least prove nodeless solutions exist in this limit \cite{BW2008_sun_ex}. Since the purely magnetic solutions are a limiting case of our dyonic solutions, we expect these results to carry over to at least some dyonic solutions, i.e. those with the EGFs all small. Hence we do not attempt to prove that \textit{all} solutions are nodeless in the MGFs as $|\Lambda|\rar\infty$; rather, we attempt to prove that we can find at least some such solutions, i.e. that this a sufficient condition for the existence of nodeless solutions. We finally note that the shrinking parameter space for fixed $\Lambda$ and increasing $N$ implies a possible necessary condition for nodeless existence involving the initial gauge field parameters, though this is something we have not yet been able to identify.

Therefore, motivated by the above results and by analogous analytical results for existence in this regime \cite{BW2008_sun_ex, JEB2014_top_ex}, we now describe transforms of our field equations that will allow us to take the limit $|\Lambda|\rar\infty$ (i.e. $\ell\rar0$) sensibly. We emphasise that we cannot let $\ell=0$ as then the asymptotic parameter $\tau$ becomes non-regular. Here, it is more convenient to take black hole and soliton solutions separately.

\begin{prop}\label{lambda0}
There exist non-trivial dyonic solutions to the field equations (\ref{EE1}, \ref{EE2}, \ref{YM1E}, \ref{YM2E}), analytic in some neighbourhood of $\ell=0$, for any choice of initial MGF values $\omega_{jh}$ (or $\breve{\beta}_{k}$) and for initial EGF values in some neighbourhood of $\mc{E}'_{jh}=0\,\,\forall j$ (or $\breve{\theta}_{k}=0\,\,\forall k$).
\end{prop}
\vsp
\textbf{Proof} We begin by considering black holes. In the purely magnetic topological case, we were able to deduce the existence and uniqueness of black hole solutions as $\ell\rar0$, by using the following rescaling to simplify the equations:
\begin{equation}\label{l0tran}
\begin{array}{lcl}
\breve{m}\equiv m\ell^2, & \quad & \breve{\mu}\equiv\mu\ell^2=k\ell^2-\disfrac{2\breve{m}}{r}+r^2.\\
\end{array}
\end{equation}
Unfortunately, in our case this does not simplify the equations much at all: all we can prove is that $\breve{m}$ must once again be a constant. However, we know that for the $\sun$ purely magnetic equations, we obtain the following unique solution (for all $r\geq r_h$):
\begin{equation}\label{l0sol}
\begin{array}{lclcl}
\breve{m}(r)=\frac{1}{3}r_h^3, & \quad & S(r)\equiv1, & \quad & \omega_j(r)\equiv\omega_{jh}.\\
\end{array}
\end{equation}

This is valid for all initial MGF values $\omega_{jh}$. Therefore, if we append $\mc{E}_j(r)\equiv0$ to this list, we will end up with a (trivially) dyonic solution as $\ell\rar0$. (Due to non-linearity it is possible that there are other solutions for $\mc{E}_j(r)\not\equiv 0$, i.e. that this solution is not unique, however we at least see that \eqref{l0sol} solves the transformed equations if and only if $\mc{E}_{j}\equiv0$.) We recall that $\mc{E}_j\equiv0\,\,\forall r$ plus the zero-sum of the $\alpha_j$ implies that $\alpha_j\equiv0\,\,\forall r$ also.

It is then also clear that the argument in Proposition 11 in previous work \cite{BW2008_sun_ex} will carry identically across for this solution, thus we have a `trivial' solution to work with for any $\ell$, including $\ell$ arbitrarily small. Hence we shall take this solution and show that by fixing $r_h$ and $\omega_{jh}$, letting $\ell$ be arbitrarily small and varying only $\mc{E}'_{jh}$, we can find solutions for non-identically-zero EGFs.

It is clear from the definition of $\mu$ that $\ell$ only makes a difference as $r$ becomes large, and is ignorable otherwise. Therefore, if we can prove local existence for $\ell\rar0$ as $r\rar\infty$ by adapting Proposition \ref{lexinf}, then Proposition \ref{globex} can be used to prove nearby non-trivial solutions exist in some neighbourhood of these purely magnetic solutions (i.e. for $\mc{E}_j$, $\ell$ small but non-zero).

We apply \eqref{l0tran} to Proposition \ref{lexinf}, and define the following new quantities:
\begin{equation}
\begin{array}{lcl}
\breve{\lambda}_1\equiv\ell^2\lambda_1, & \quad & \breve{\mc{F}}_\infty\equiv\ell^2\mc{F}_\infty=\ell^2\mu x^2.\\
\end{array}
\end{equation}
Examining the proof of Proposition \ref{lexinf}, we see that all of the polynomials $\mc{G}_{\infty,2}$, $\mc{H}_{\infty,1j}$, $\mc{H}_{\infty,2j}$ change internally but remain regular (including as $\ell\rar0$). We must also replace the equation $x(d\Lambda/dx)=0$ with
\begin{equation}
x\frac{d\ell}{dx}=0;
\end{equation}
but the only equation which changes is that of $\lambda_1$, which becomes
\begin{equation}
x\frac{d\breve{\lambda}_1}{dx}=x\breve{\mc{G}}_{\infty,1}\equiv x\ell^2\mc{G}_{\infty,1}.
\end{equation}
Therefore we see that this proposition remains valid as $\ell\rar0$. As for the asymptotic regime, we can see that our argument in Section \ref{asymptotics} remains valid for all $\ell$ small but non-zero.  

The argument that proves the existence of black hole solutions in the regime is essentially identical to that in Proposition \ref{globex}. We fix $r_h$ and $\omega_{j,h}$ non-zero and we fix $\ell$ arbitrarily small. We then choose some $r_1 >> r_h$ so that we can consider $r_1$ in the asymptotic regime, and we consider varying only $\mc{E}'_{j,h}$. Then Propositions \ref{lexrh} and \ref{gloreg} confirm that for $\mc{E}'_{j,h}$ sufficiently small we can find solutions near purely magnetic solutions which will begin near $r=r_h$ and remain regular in the range $(r_h,r_1]$, and that those solutions will be nodeless in the MGFs due to analyticity. Finally, once we are in the asymptotic regime, we can use the logic in Section \ref{asymptotics} to ensure that solutions will remain regular as $r\rar\infty$ and that all $\omega_j$ will have no nodes.

Luckily in the case of solitons, the argument is very similar. The analogous argument for existence of purely magnetic solutions in this regime exists already \cite{BW2008_sun_ex}, so we will just mention that we must be more careful about how we take the limit $\ell\rar0$ since we must use the parameter $r^{-1}$ at infinity. For black holes this is fine since $r_h>0$ and therefore $r^{-1}$ is bounded and thus we have a natural `scale' to work with, but for solitons, we must have $r=0$ at some point so that $r^{-1}$ is not bounded. 

There, we defined the following transforms,
\begin{equation}
r=\ell\tilde{x}, \quad\quad\quad m(r)=\ell\breve{m}(\tilde{x}),
\end{equation}
and found that for the MGFs, it was easier to work with functions defined by the basis of eigenvectors that we used near the origin:
\begin{equation}
\bsym{\omega}=\bsym{\omega}_0+\slim_{k=2}^N\varpi_k(r)\bsym{v}_kr^k=\bsym{\omega}_0+\slim_{k=2}^N\varpi_k(\ell\tilde{x})\ell^k\tilde{x}^k\bsym{v}_k.
\end{equation}
In that case, the solution turns out to be
\begin{equation}\label{l0sol1}
\begin{array}{lclcl}
\breve{m}(r)\equiv0, & \quad & S(r)\equiv1, & \quad & \varpi_k\propto\,_2F_1\left(\disfrac{k+1}{2},\disfrac{k}{2};\frac{2k+1}{2};-x^2\right).\\
\end{array}
\end{equation}
The functions $\varpi_k$ are essentially the same as $\bar{\beta}_k$ but for a factor of $\gamma_j$. It can be noted that the properties of hypergeometric functions can be used to establish that \eqref{l0sol1} also fulfils the boundary conditions at $r\rar\infty$. Using a very similar argument to the black hole case, we append $\mc{E}_j\equiv0\,\,\forall j$, i.e. $\bar{\theta}_k(x)\equiv0\,\,\forall k$, to our solution \eqref{l0sol1}; and we deduce that we can find global solutions regular and analytic in all field variables for $\ell$ arbitrarily small, for arbitrary values of $\breve{\beta}_k$ and for $\breve{\theta}_k$ small ($\forall k$).$\Box$

\section{Conclusions}
\label{conclu}

The purpose of this research was to investigate the existence of black hole and soliton solutions to topological adS dyonic equations in four-dimensional $\sun$ EYM theory, motivated by previous existence results for purely magnetic spherical $\sun$ solutions \cite{BW2008_sun_ex}, purely magnetic topological $\sun$ black hole solutions \cite{JEB2014_top_ex}, and dyonic $\essu$ spherical solutions \cite{NW2012_dyon}.

We began by using a previously derived gauge potential appropriate to the case \cite{JEB2014_top_ex}. We used this to derive the field equations in this case -- two Einstein equations and $2N-2$ independent Yang-Mills equations -- and found several trivial solutions, including an embedding of $\essu$ in $\sun$ (Proposition \ref{embedprop}). At $r=r_h$ and $r\rar\infty$, we used physicality requirements to establish appropriate boundary conditions in an elementary fashion. For solutions regular at the origin, the situation was much more complicated and required detailed analysis.

Upon expanding the equations in power series at the origin, we discovered that to establish consistency of solutions, we needed to solve a tri-diagonal system similar to one previously considered by K\"{u}nzle \cite{HPK1994_analysis_sun}. In our case we have a mutual system of equations, but we found that each set may be expanded in the same basis of eigenvectors, which simplified things a little. We used a form of `ladder' induction over two induction sentences to prove that each expansion parameter in each system depended only on previous parameters, and that for each gauge field we needed $N-1$ independent parameters to describe the power series at $r=0$ (Proposition \ref{conprop}).

We then proceeded to prove the existence of solutions locally near the boundaries (Propositions \ref{lex0}, \ref{lexrh} and \ref{lexinf}), which are regular and analytic in their boundary values, using an established theorem of ordinary differential equations (Theorem \ref{exdetheor}). Existence is elementary to establish in the cases of $r=r_h$ and $r\rar\infty$, and a little more complicated for regular solutions at $r=0$. Nonetheless we establish local existence here too.

After that, we constructed a series of arguments which form the heart of the proof. After proving that all $\mc{E}_j$ are monotonic (Proposition \ref{emon}) we proved that if a solution remains regular over a small interval near the event horizon (origin), then with the condition $\mu(r)>0$ for all $r>r_h$ ($r>0$) we may continue to integrate that solution out regularly into the asymptotic regime (Proposition \ref{gloreg}). We examined this regime and discovered that due to the parameter we used to render the asymptotic equations autonomous, the solution would continue to remain regular for $r$ arbitrarily large (Section \ref{asymptotics}). Finally, we proved that for these field equations, solutions exist in open sets (Proposition \ref{globex}); and given that the literature is now fairly abundant with known solutions that result from trivialising these field equations in some way \cite{VdBR_su2_top, BW2008_sun_ex, JEB2014_top_ex, NW2012_dyon}, we are able to prove the global existence of non-trivial solutions in a number of regimes (Theorem \ref{regimes}). In addition, we proved that solutions could be found in the limit $|\Lambda|\rar\infty$ for arbitrary initial parameters for $\omega_j$, and for initial parameters of $\mc{E}_j$ small (Proposition \ref{lambda0}).

The main result of this paper is the proof of existence of nodeless non-trivial black hole and soliton solutions to four-dimensional topological $\sun$ dyonic EYM equations in various regimes: nearby existing solutions, and in the limit $|\Lambda|$ large. In particular, we have shown that we may dress a black hole or soliton with an arbitrarily large amount of gauge field hair. There are several future research directions that are suggested by the results here. The possibility of dressing a dyonic black hole with arbitrary amounts of hair suggests there may be some work to do in extending the gauge group as large as we possibly can, and therefore the dyonic $\suinf$ case becomes interesting. We note that the purely magnetic case has already been considered \cite{EW_suinf_1999}, in which evidence of the existence of solutions is provided. String theories are characterised by enormous gauge symmetries and hence provide a motivation for this work.

Another possible extension we could make is suggested by Gubser \cite{gubser_colorful_2008}, who found some very interesting results for dyonic $\essu$ planar black holes in adS; namely, a second-order phase transition between the embedded planar RNTadS black hole and a black hole with a non-trivial Yang-Mills field condensate. It would be natural to ask how his results generalise to $\sun$ in light of the solutions that we have discovered here.

In addition to this there are questions which arise about the impact of black hole hair on other areas of gravitational physics. For instance, it may be valuable to consider the adS/CFT (Conformal Field Theory) correspondence in light of this work \cite{maldacena_large_1998, witten_anti_1998, witten_anti-sitter_1998}, since it has been conjectured that there are observables in the dual CFT which are sensitive to the presence of black hole hair \cite{hertog_black_2004, gubser_colorful_2008}. It would also be of interest to know whether these topological models would be valuable to modelling holographic superconductors, since planar black hole models have recently been used in this research area \cite{cai_introduction_2015}.

Finally, there is the important question of further confirming or refining the generalised `No-hair' theorem, on which this work will have direct implications. The statement of this theorem given by Bizon \cite{PB1990_su2_BH} is:
\begin{quote}
``In any given matter model, \textit{stable} black holes will be characterised by a finite number of global charges.'' 
\end{quote}
Therefore, the next important problem here is to examine the stability of the solutions that we have found, since we can establish the existence of stable solutions in the purely magnetic spherical case \cite{EB+EW_stab_sun}, and results have just emerged proving the stability of some topological black hole solutions \cite{JEB_WIP_stab_top} and $\essu$ dyonic solutions \cite{BN+EW_su2_dyon_stab}. We expect however that establishing the stability of our dyonic solutions will be a highly non-trivial problem, since the presence of the electric gauge field will disallow the decoupling that happens in the purely magnetic case, making it much harder to prove analytical assertions.

%



\begin{thebibliography}{61}
\bibitem [{foo()}]{footnote1}%
\BibitemOpen
\href@noop {} {\bibinfo  {journal} {Note that we shall mostly refer to such
		solutions as `solitons' in this work due to the generality of the term}\
}\BibitemShut {NoStop}%
\end{thebibliography}

%

\end{document}